  \documentclass[10pt,prb,nobibnotes,showpacs,showkeys,superscriptaddress,maxnames=6]{revtex4-1}
\usepackage{amsfonts,amsmath,amssymb,graphicx, placeins, bm}
\usepackage{setspace,color,cases}
\usepackage{pstricks}
\usepackage{fdsymbol}

\usepackage{ae,aeguill}
\usepackage[utf8]{inputenc}    
\usepackage[T1]{fontenc}
\newcommand{\nna}{\hat{n}_i\hat{n}_j -3 (\hat{n}_i \hat{r}_{ij})(\hat{n}_j \hat{r}_{ij})}

  \newcommand{\be}{\beta}     
\newcommand{\la}{\lambda} \newcommand{\al}{\alpha}   
  \newcommand{\ep}{\epsilon} \newcommand{\s}{\sigma}
     
 \newcommand{\tend}{\rightarrow}
\newcommand{\equa}[1]{\begin{eqnarray} \label{#1}}
\newcommand{\auqe}{\end{eqnarray}}
\newcommand{\equab}[1]{\begin{widetext}\begin{eqnarray} \label{#1}}
\newcommand{\auqeb}{\end{eqnarray}\end{widetext}}
\newcommand{\tab}[1]{\begin{tabular}{#1}}
\newcommand{\bat}{\end{tabular} \\ }
\begin {document}
\title 
 {Phase diagram of a three-dimensional dipolar Ising model with textured Ising axes.}

\author{V. Russier}
\email[e-mail address: ] {russier@icmpe.cnrs.fr}
\affiliation{ICMPE, UMR 7182 CNRS and UPE 2-8 rue Henri Dunant 94320 Thiais, France.}
\author{Juan J. Alonso}
\email[e-mail address: ] {jjalonso@uma.es}
\affiliation{F\'{\i}sica Aplicada I, Universidad de M\'alaga, 29071 M\'alaga, Spain}
\affiliation{Instituto Carlos I de F\'{\i}sica Te\'orica y Computacional,  Universidad de M\'alaga, 29071 M\' alaga, Spain}

\date{\today}   
\pacs{75.10.Hk, 75.10.Nr, 75.40.Cx, 75.50.Lk}

\begin {singlespace}
\begin{abstract}
We study from tempered Monte Carlo simulations the magnetic phase diagram of a textured dipolar Ising model on a 
face centered cubic lattice. The Ising coupling of the model follow the dipole-dipole interaction.
The Ising axes are distributed with a uniaxial symmetry along the $\hat{z}$ direction with a gaussian probability 
density of the polar angles. This distribution provides a quenched disorder realization of the dipolar Ising
model making a continuous link between the parallel axes dipoles and the random axes dipole models.
As expected the phase diagram presents three distinctive phases: paramagnetic, ferromagnetic and spin-glass.
A quasi long range ferromagnetic and a reentrant spin-glass phases are obtained in the vicinity of the ferromagnetic 
spin-glass line. This model provides a way to predict the magnetic phases of magnetic nanoparticles supracrystals
in terms of the texturation of the easy axes distribution in the strong anisotropy limit.

\end{abstract}
\end {singlespace}
\maketitle
%
 \section {INTRODUCTION}
 \label {intro}

Dipolar Ising models present a rich variety of ordered phases in 3 dimensions,
including ferromagnetic (FM), antiferromagnetic (AFM), paramagnetic (PM) and glassy phases 
according to the disorder and frustration stemming from the underlying structure and/or dilution
or additional short range exchange interaction.
This diversity results from the long range dipole-dipole interaction (DDI) 
whose anisotropy leads to both a ferromagnetic and an anti-ferromagnetic couplings
and is the driving force of the collective effects. Dipolar Ising models (DIM) are particularly 
suitable to model dipolar crystals~\cite{reich_1990,biltmo_2009}. 
They are also well adapted to model the magnetic phases of single domain magnetic 
nanoparticles (MNP) assembled in densely packed configurations~\cite{bedanta_2009,bedanta_2013} 
at least in the limit of strong uniaxial anisotropy. 
These latter systems are the focus of a large activity in the field of nanoparticle research 
because of their wide range of potential applications and because they 
provide convenient experimental samples for the study of nanoscale magnetism. 
Of particular interest are the ensembles of MNP self assembled in superlattices 
(or supracrystals), $i.e.$ ordered crystals made of MNP whith long range 
order~\cite{lisiecki_2003b,lisiecki_2012,mishra_2012,mishra_2014,josten_2017,ngo_2019}.
When the considered MNP ensemble are made of spherical MNP with a sharp size distribution, 
and coated by a non magnetic layer preventing aggregation, the structure of the resulting 
ordered crystal, following mainly the rules of hard sphere packing is in general of BCT, 
FCC or HCP symmetry~\cite{boles_2016}. 
An important experimental situation is thus that of MNP self organized in supracrystals with 
FCC~\cite{lisiecki_2003b,lisiecki_2012,josten_2017,ngo_2019} or HCP~\cite{mishra_2014} symmetry.
This leads to an ordered lattice with close packing symmetry of nanoparticles 
which, owing to the non magnetic coating layer, 
behave as dipoles undergoing an effective anisotropy which drives the dipole moment
toward the easy axis of magnetization.
As a model for MNP assemblies, the dipolar Ising model where the direction of the dipoles are 
imposed on the easy axes corresponds strictly speaking to the limit of infinite value of 
the effective magnetocrystalline anisotropy energy (MAE) with uniaxial symmetry. 
Such a limit is reasonable in the light of actual experimental
situations, where one expects typically the MAE of an order of magnitude larger than the 
DDI contribution.
We emphasize that the MAE being a one-body potential does not couple the moments and
is to be considered in the framework of the blocking process which 
involves the measuring time~\cite{dormann_1997,skomski_2003}. 
Hence the relevant criterion to determine whether the DDI play an important role is the ratio of
the characteristic DDI temperature ($\sim{}E_{dd}/k_B$) to the blocking temperature $T_b$ of the dispersed system,
$E_{dd}$ being the DDI energy per particle taking into account the MNP concentration. 
Strictly speaking the dipolar Ising model is to be considered as the infinite measuring time or equivalently the 
$T_b\tend{}0$ limit.
For magnetometry measurements, $T_b\simeq{}E_a/30k_B$~\cite{dormann_1997,bedanta_2009}, 
where $E_a$ is the MNP anisotropy barrier, leading typically to $T_d/T_b\gtrsim{}5$
which gives sense to the dipolar Ising model in such cases.
In addition to the well ordered lattice, one can also consider in the framework of DIM 
the random close packed structure of hard spheres~\cite{alonso_2017,alonso_2019}, as ensembles of 
MNP presenting such a structure can be obtained experimentally from sintered powders and were analyzed 
on the framework of spin glass behavior~\cite{de-toro_2013a,de-toro_2013b,andersson_2017}. 

Whatever the structure of the MNP ensemble, the high temperature magnetic phase is paramagnetic
in nature (the so-called superparamagnetic regime) and one key question remains to determine
and to predict the nature of the low temperature ordered phase in highly concentrated systems 
where collective effects are expected. The amount of disorder is obviously a crucial parameter. 
When dealing with ordered supracrystals, the easy axes distribution, $\{\hat{n}_i\}$, plays 
then a central role. 
For a colloidal crystal synthesized in the  absence of external field, $\{\hat{n}_i\}$ 
is a random distribution while in the case of a synthesis under external field one expects 
the possibility to get a textured distribution of easy axes along the direction of the field 
before freezing.

The general experimental finding for the self organized, or compact assemblies of MNP 
in the absence of texturation is a spin glass
frozen phase which can be understood both by the strong anisotropy and the random
distribution of easy axes. 
In Ref.~\cite{nakamae_2010} the easy axes alignment has been obtained in a frozen ferrofluid
via the external field during the freezing  of the embedding non magnetic matrix. However,
the lack of structural order and the low MNP concentration leads
also to a spin glass state at low temperature. 

In order to model the above situations, we must consider the easy axis distribution as 
the relevant parameter controlling the amount of disorder in the system. 
While the onset of ordered phase for concentrated dipolar systems free of MAE (either FM
or AFM) is well documented~\cite{luttinger_1946,wei_1992a,weis_1993,bouchaud_1993,weis_2005a},
there is a lack of knowledge on the influence of the easy axes texturation on both the 
nature of the ordered phase and the value of the corresponding transition temperature. 
The dipolar Ising model with random distribution of Ising axes, the random 
axes dipoles model (RAD)~\cite{fernandez_2009,alonso_2017}, 
is known to present a spin-glass (SG) ordered phase at low temperature.
On an other hand, the totally oriented dipolar Ising model, parallel axes dipoles model 
(PAD)~\cite{klopper_2006,fernandez_2000,alonso_2010,alonso_2015} 
presents a long range ferromagnetic (or antiferromagnetic for the simple cubic lattice) phase when 
the concentration of occupied sites takes a value larger than a threshold one
($x_c\simeq{}0.65$ on the simple cubic lattice).
Klopper {\it et al.}~\cite{klopper_2006} have considered a dipolar Ising model on a FCC lattice
with a random exchange term as source of disorder. 
Their result is an ordered FM phase for small enough values of the random exchange coupling
$J_{EA}$ with a strongly $J_{EA}$ dependent PM/FM transition temperature.
In Ref.~\cite{alonso_2019} both the easy axes texture and particles structure through a random close
packed  distribution are considered as sources of disorder.

In this work we investigate, through Monte Carlo simulations, the dipolar Ising model on a 
perfect FCC lattice with textured easy axes distribution which we denote by the 
textured axes dipoles model (TAD). 
The FCC lattice is chosen first as a convenient example for spontaneous dipolar 
ferromagnetic order and, as mentioned above, for its relevance for experimental situations.
Our purpose is to investigate the magnetic phase diagram in terms of the variance $\s$ 
of the polar angles distribution of the Ising axes relative to the $\hat{z}$ axis. 
The ordered phases in the limiting cases $\s=0$ (PAD) and $\s\tend\infty$ (RAD) are ferromagnetic and 
spin-glass respectively and we thus focus on the determination of the value of $\s$ 
corresponding to the FM/SG line on the one hand and on the determination of $T_c(\s)$ 
along the PM/FM and PM/SG lines on the other hand. In case of the ferromagnetic ordering 
at low temperature, we also have to characterize the phase according to its long range versus 
quasi long range order.

The rest of paper is organized as follows. We first describe the model in section~\ref{model}, 
then we give some indications on the Monte Carlo scheme and present the observables we focus on. 
Section~\ref{results} is devoted to the analysis of our results and we conclude in section~\ref{concl}.
 \section {Model}
 \label   {model}
 We consider a system of dipoles of moment $\mu$ located on the sites of a perfect face centered cubic (FCC) 
lattice, interacting through the usual dipole dipole interaction (DDI) and constrained to point along the easy axes, $\hat{n}_i$, 
defined on each lattice site. 
The distribution of easy axes, $\{\hat{n}_{i}\}$ is characterized by the texturation in the direction $\hat{z}$ with axial symmetry.
To this aim, the azimuthal angles are randomly chosen while the polar angle $\{\Theta_i\}$ distribution follow the probability density
\equab{p_theta}
 p(\Theta) = C sin(\Theta)  \left[ exp(-(\Theta^2/2\s^2) + exp(-((\pi-\Theta)^2/2\s^2))\right]
\auqeb
where $C$ is a normalization constant and $sin(\Theta)/2$ corresponds to the random distribution. 
The variance $\s$ of this probability distribution is to be considered as the disorder control parameter of our model as the 
dilution in the random diluted Ising model or the short range random exchange term $J_{EA}$ in 
the dipolar Ising plus random exchange model of Ref.~\cite{klopper_2006}.
$\s=0$ obviously corresponds to the totally aligned or textured model (PAD) while $\s\tend\infty$ corresponds to the random
distribution of the $\{\hat{n}_i\}$ (RAD). Practically we find that the actual limit of $\s$ beyond which the random distribution 
is reached is merely $\s_s\simeq{}\pi/2$. The hamiltonian of the system is given by
\begin{subequations}\label{edd_1}
\equab{edd_n}
  \be H & = & 
        \frac{1}{2}\frac{\be}{\be_0} \ep_d \sum_{i\neq j} s_i{}s_j \frac{\nna}{(r_{ij}/d)^3}
        ~~~ \textrm{with}~ \ep_d = \be_0\frac{\mu_0}{4\pi}\frac{\mu^2}{d^3} 
        \label{edd_1a} \\
  ~ & \equiv & ~ 
        \frac{1}{2}\frac{\be}{\be_0} \ep_d \sum_{i\neq j} s_i{}s_j \frac{D_{ij}}{(r_{ij}/d)^3
        \label{edd_1b}}
\auqeb
\end{subequations}
where $s_i=\pm{}1$ is the set of Ising variables, related to the dipole moments $\vec{\mu}_i = \mu{}s_i\hat{n}_i\equiv{}\mu\hat{\mu}_i$, 
$\hat{r}_{ij}$ is the unit vector carried 
by the vector joining sites $i$ and $j$, $\be=1/(k_BT)$ is the inverse temperature, $T_0$ ($1/k_B\be_0$) a characteristic 
temperature of the actual system. For instance $T_0$ can be chosen in such a way that $\ep_d=1$ and the same model can
represent different systems according to $T_0=(1/k_B)(\mu_0/4\pi)\mu^2/d^3$.
$d$ is a unit of length, chosen as the nearest neighbor distance between dipoles, here that of the FCC lattice,
or alternatively the nanoparticle diameter, $d_p$, when the model is applied to a MNP ensemble. 
Concerning the reduced temperature, instead of the natural choice $T/(T_0\ep_d)$, we take advantage of the $1/r^3$ 
dependence of the DDI
and of the properties of the sum entering in equation~(\ref{edd_1}), see note~\cite{note_t*}, 
to introduce the more convenient reduced temperature $T^*=T/(T_0\ep_d(\Phi/\Phi_r))$,
where $\Phi$ and $\Phi_r$ are either the number of occupied sites per unit volume and a reference value (here the 
number of sites per unit volume of the FCC lattice) or alternatively the MNP volume fraction ($(N/V)\pi{}d_p^3/6$) and a 
reference value (for instance, the maximum value for hard spheres on a FCC lattice).
This reduced temperature makes easier the comparison of results of systems closely related but presenting different volume fractions and/or
structure, as will be discussed in section~\ref{concl} when comparing the phase diagrams of the FCC lattice and the RCP cases. 

The simulation box is a cube with edge along the $\hat{z}$ direction and edge length $L_s=\sqrt{2}Ld$ and the total number of 
dipoles is $N=4L^3$. The close packed direction of the FCC lattice is (1,1,1).
We consider periodic boundary conditions by repeating the simulation cubic box identically in the 3 dimensions.
The long range DDI interaction is treated through the Ewald summation technique~\cite{allen_1987,weis_1993}, 
with a cut-off $k_c=10k_m$, $k_m=(2\pi/L_s)$, in the sum of reciprocal space and the $\al$ parameter of the 
direct sum is chosen either $\al=5.80$ or 7.80~\cite{weis_1993}.
In such conditions the errors introduced by the periodic boundary conditions in the framework of the Ewald
summation technique are known to be very small even at low temperature and to vanish in the thermodynamic
limit~\cite{wang_2001}.
The Ewald sums are performed with the so-called 
conductive external conditions~\cite{allen_1987,weis_1993}, i.e. the system is embedded in a medium with 
infinite permeability, $\mu_s=\infty$, which is a way to avoid the demagnetizing effect and thus 
to simulate the intrinsic bulk material properties regardless of the external surface and system 
shape effects. 
The equivalence between simulation results for the system embedded in vacuum, $\mu_s=1$, or in 
the infinite permeability medium, $\mu_s=\infty$ can be done through the introduction of an 
external field and noting that the relevant field is either the internal one if $\mu_s=\infty$ or the 
external one if $\mu_s=1$~\cite{russier_2013}.

\subsection {Simulation method}
\label{sim_meth}

In order to thermalize in an efficient way our system presenting strongly frustrated states,
we use parallel tempering algorithm~\cite{earl_2005} (also called tempered Monte Carlo) for our 
Monte Carlo simulations. Such a scheme is widely used in similar systems, and we do not enter in 
the details. The method is based on the simultaneous simulation runs of identical replica 
of a system with a given distribution of axes $\{\hat{n}_i\}$
for a set of temperatures $\{T^*_n\}$ with exchange trials of the configurations pertaining to different 
temperatures each $N_M$ Metropolis steps according
to an exchange rule satisfying the detailed balance condition. The set of temperatures is chosen
in such a way that on the first hand it brackets the paramagnetic ordered state transition temperature
and on an other hand it leads to a satisfying rate of exchange between adjacent temperature configurations.
Our set $\{T^*_n\}$ is either an arithmetic distribution or 
an optimized one in order to make the transfer rate between adjacent paths as constant as possible
in the whole range of $\{T^*_n\}$. We obtained close behavior for the rate of transfer by using the 
efficient constant entropy increase~\cite{sabo_2008} or the simpler geometrical distribution of 
temperatures. In the present work we take $N_M=10$, 
the number of temperatures is in between 36 and 60 according to the value of $\s$ and the amplitude
of temperatures in the set $\{T^*_n\}$. The minimum and maximum values of the set $\{T^*_n\}$ 
depends on $\s$ when dealing with the PM/FM line, as does the critical temperature 
$T^*_c$ (see figure~(\ref{diag_phase_tad})). 
The simulations for $\s\ge{}0.60$ are mainly performed with $n=60$, $T^*_n\in{}[0.55,3.50]$ 
and an arithmetic distribution.

When necessary, precise interpolation for temperatures between the points actually 
simulated are done through reweighting methods~\cite{ferrenberg_1988}.
We use $t_0$ Monte Carlo steps (MCS) for the thermalization and  the averaging is
performed over the $(t_0,2t_0)$ following MCS with $t_0=5\;10^5$ for $\s\in[0,0.5]$ namely sufficiently 
away from the FM/SG line and $t_0=10^6$ otherwise.

We deal with frozen disorder situations where each realization of the easy axes distribution 
$\{\hat{n}_i\}$ defines a sample. Accordingly, a double averaging process is performed first relative 
to the thermal activation, the Monte Carlo step, and second on the whole set of $N_s$ samples.
Consequently, the mean value of an observable $A$, results from a double averaging denoted in the following as
$[<A>]$ where $<.>$ corresponds to the thermal average on the MC sampling for a fixed realization of
the axes distribution and $[.]$ to the average over the set of samples considered.
The number of samples necessary to get an accurate result depends strongly on the value of $\s$.
Obviously, for $\s=0$, $N_s=1$ should be sufficient for a very long MC run in order to get a satisfying
average. Practically, we find $N_s$ of the order of 300 to 500 sufficient up to $\s=0.4$ while 
when $\s$ gets closer to the value corresponding to location of the SG/FM line up to $N_s=12\;000$ realizations
are necessary.

\subsection {Observables}
\label {obs}

Our main purpose is the determination of the transition temperature between the paramagnetic and the ordered 
phase and on the nature of the latter, namely ferromagnetic or spin-glass, in terms of $\s$. 
For the PM/FM transition, 
we consider the spontaneous magnetization 
\equa{m_1}
  m = \frac{1}{N} \left\Arrowvert \sum_i \hat{\mu}_i \right\Arrowvert
\auqe
computing its moments, $[<m^k>]$, k = 1,2 and 4.
We compute also the nematic order parameter $\la$ together with the instantaneous nematic direction, $\hat{d}$ which are 
the largest eigenvalue and the corresponding eigenvector respectively of the tensor
$  
 \bar{Q} = \frac{1}{N} \sum_i (3\hat{\mu}_i\hat{\mu}_i - \bar{I})/2.
$ 
$\bar{Q}$ is the orientational tensor and $\la$ the related second rank order parameter of nematic 
liquids theory~\cite{vieillard-baron_1974} and widely used in studies of dipolar 
systems~\cite{wei_1992a,weis_1993,chamati_2016,russier_2017}, the value of the latter being expected to range 
from $\la=0$ and 1.0 in the random and totally oriented cases respectively.
Conversely to the true dipolar system, in the dipolar Ising model, the $\hat{\mu}_i$ can be replaced by the $\hat{n}_i$ in
$\bar{Q}$ and both $\la$ and $\hat{d}$ become characteristic of the easy axes distribution of each sample 
and accordingly are first $T^*$-independent and second totally determined by $\s$ in the limit $N\tend\infty$
(obviously $\la=1$ for $\s=0$, otherwise we get $\la\simeq{}0$ for $\s\ge\pi/2$ and for instance $\la=0.630$ for $\s=0.40$).
The spontaneous magnetization can also be studied in the ordered phase from the 
projected total magnetization on the nematic direction~\cite{weis_1993}, which defines 
\equa{m_la}
   m_{\la} =  \frac{1}{N} \sum_i \hat{\mu}_i.\hat{d} 
\auqe 
Given the axial symmetry along $\hat{z}$ in the present model, when $\s{}\le0.8$ we find  
that the sample to sample fluctuations of $\hat{d}$ around $\hat{z}$ are vanishingly small and $m_{\la}$ is accordingly 
nearly indistinguishable from the $z-$component $<m_z>$ and therefore $<m_{\la}>$ and $<m_z>$ play the same role. 
We compute the the mean $m_1=[<|m_{\la}|>]$ and the moments $m_{n}=[<m_{\la}^n>]$, with $n=2,4$.
To locate the transition temperature, $T_c$, as usually done, we will use the the finite size scaling (FSS) analysis of the Binder
cumulant which is defined on the component $m_{\la}$
 \equa{bm_la}
   B_m = \frac{1}{2} \left( 3 - \frac{m_{4}}{m_2^2} \right)
 \auqe
From this normalization, $B_m\tend{}1$ in the long range FM phase and $B_m\tend{}0$ in the limit $L\tend\infty$ in the disordered 
PM phase. For the PM/SG transition, we consider the usual overlap order parameter
\equa{q}
  q = \frac{1}{N} \sum_i s_i^{(1)}s_i^{(2)}
\auqe
where the superscripts $^{(1)}$ and $^{(2)}$ stand for two identical and independent replicas of the same sample. From $q$ 
we calculate the mean value $[<|q|>]$ and its moments, $q_k = [<q^k>]$ with k = 2, 4, and the corresponding Binder cumulant,
 \equa{bq}
    B_q = \frac{1}{2}(3 - \frac{q_4}{q_2^2})
 \auqe 
Finally, the magnetic susceptibility, $\chi{}_m$ and the heat capacity , $C_v$ are calculated from the magnetization and the energy 
fluctuations respectively
\equa{fluct}
  \chi_m = \frac{N}{T^*}\left( m_2 - m_1^2 \right) ~~~, \nonumber \\
   C_v = \frac{1}{NT^{*2}} \left[ <H^2> - <H>^2 \right ]
\auqe
%

 \section    {Results}
 \label {results}
\begin {figure}[h!]
  \includegraphics* [width = 0.50\textwidth , angle = -90.00]{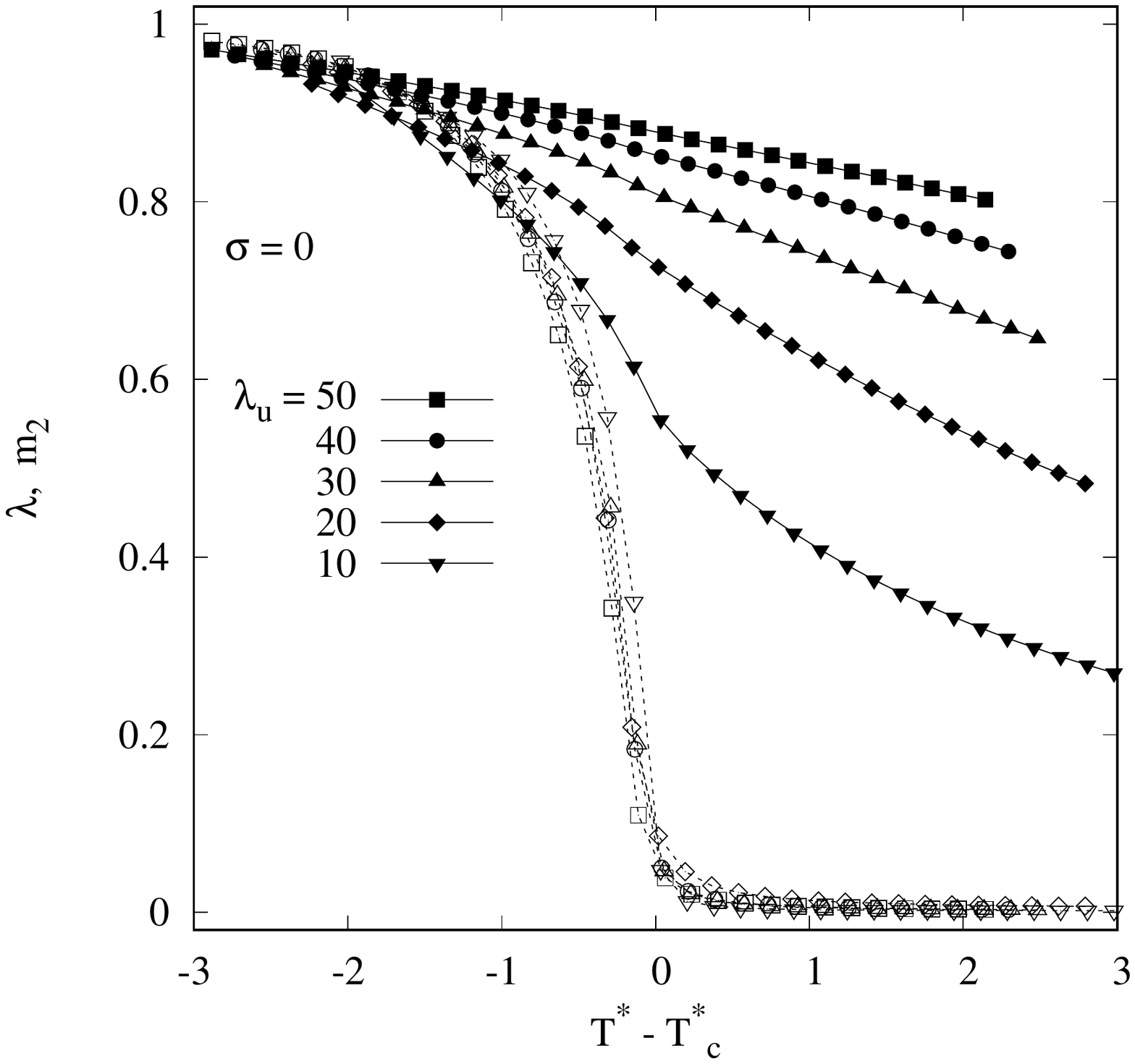}
  \caption {\label {lambda_eu}
  Nematic order parameter, $\la$ (solid symbols) and $m_2$ (open symbols) for the dipolar system for $\s=0$ and MAE coupling $\la_u$ as indicated.
  $L=6$. $T^*_c$ is estimated from the crossing point of the Binder cumulant $B_m$ with $(L_1,L_2)=(4,6)$.
  }
  \end{figure}

 We first examine whether a dipolar system with uniaxial anisotropy of finite amplitude can be, at least qualitatively,
represented by the present DIM. For this we note that the dipolar system on a FCC lattice with small or vanishing value 
of the uniaxial anisotropy present a PM/FM transition which is directly related to the onset of the nematic order measured 
from the tensor $\bar{Q}$~\cite{weis_1993,chamati_2016,russier_2017}. Conversely the nematic order in the DIM is frozen and 
determined by the Ising axes distribution as mentioned in section~\ref{model}. Therefore, we conclude on the Ising-like 
character of the PM/FM transition when the latter becomes uncorrelated from the onset of the nematic order. 
As an example we consider a totally textured system ($\s=0$).
The dipolar system is obtained by replacing in the hamiltonian~(\ref{edd_1}) the $\hat{n}_{i}s_i$ by the moments $\hat{\mu}_i$
and adding the uniaxial magnetocrystalline anisotropy energy (MAE) $\be{}E_a=-(\be/\be_0)\ep_a\sum_i(\hat{n}_{i}\ldotp\hat{\mu}_i)^2$, 
the MAE coupling constant being defined by $\la_u=\ep_a/(\ep_{d}\Phi/\Phi_r)$.
We conclude from the evolution of $\la(T^*)$ for the dipolar system across the PM/FM transition shown in figure~(\ref{lambda_eu}) that
the transition takes an Ising character for $\la_u\ge{}20$ and is still qualitatively Ising like for $\la_u\in[10,20]$. 
This means that when $\la_u\ge{}20$, although the value of $T^*_c$ depends on $\la_u$
the moments remain strongly aligned along the easy axes in the paramagnetic phase.  
 \subsection {Phase diagram}
 \label {phase_diag}
\begin {figure}[h!]
  \includegraphics* [width = 0.50\textwidth , angle = -90.00]{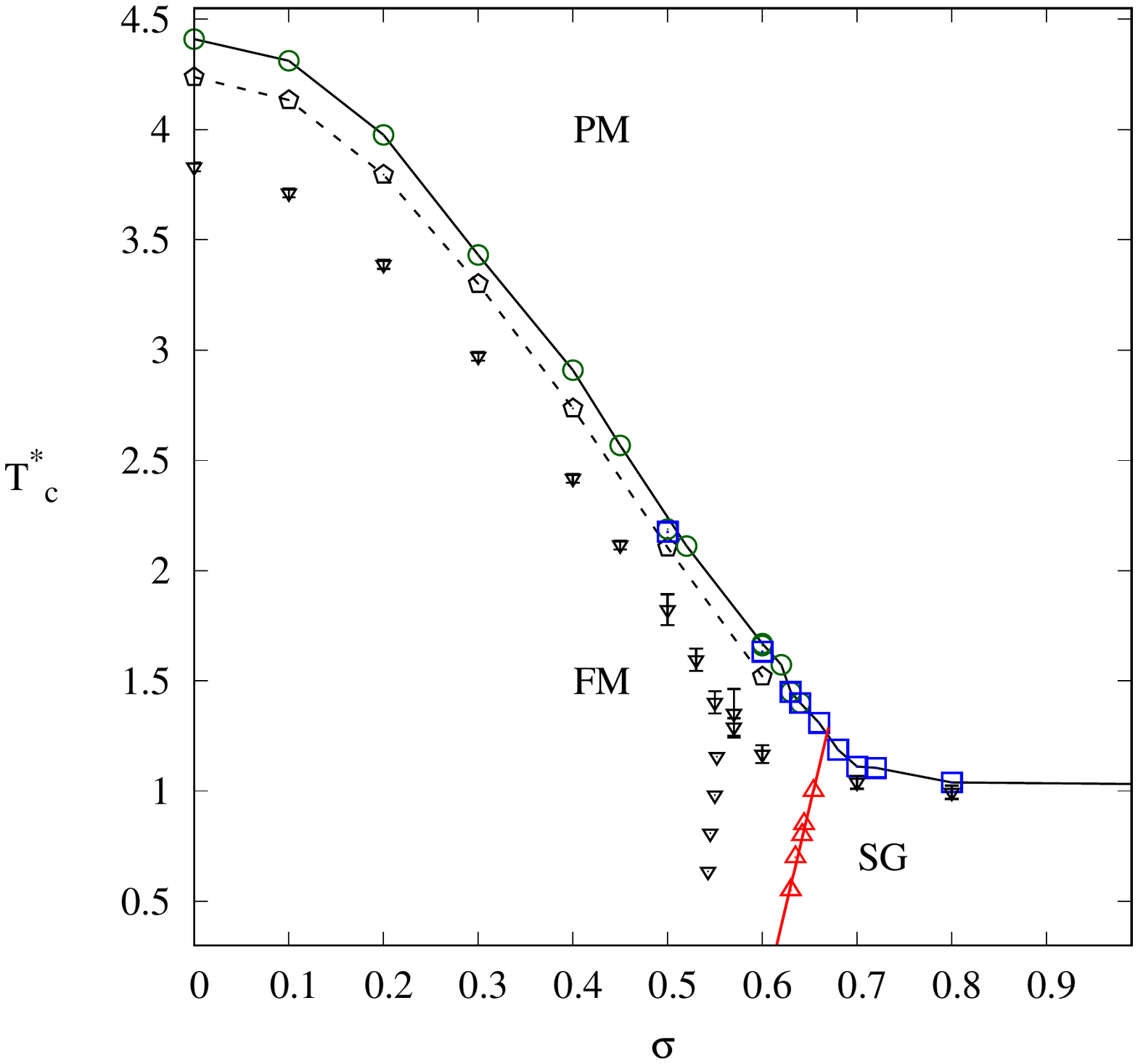}
  \caption {\label {diag_phase_tad}
  Phase diagram $T_c^*(\s)$ determined from the magnetization Binder cumulant, $B_m$, open circles;
  the overlap SG order parameter Binder cumulant $B_q$, open squares. The SG/FM line is obtained from the $L$ dependence of
  $B_m(\s)$ at fixed $T^*$, triangles. $T^*_c$ on the PM FM line obtained from the maximum of $\chi_m$ for $L=8$ ($\s=0$-0.4) 
  or $L=7$ ($\s=0.5$ and 0.6), diamonds and dashed line. Lines are guides to the eyes.
  For comparison, the transition points corresponding to the RCP case~\cite{alonso_2019} are displayed as downward triangles.
  The line beyond $\s=0.8$ is an interpolation to the value of $T_c^*$ corresponding to 
  the random distribution (RAD), see table~(\ref{tab_tc}). 
     }
  \end {figure}

 The global phase diagram in the plane $(T^*,\s)$ of our textured dipolar Ising model whose determination 
is our main objective and is given on figure~(\ref{diag_phase_tad}), separates the three distinctive phases PM, FM and SG.
The salient features are a strongly $\s$ dependent $T^*_c(\s)$ on the PM/FM line,
 a nearly constant $T^*_c(\s)$ on the PM/SG line and a weakly reentrant behavior on the FM/SG line.
 On a qualitative point of view this compares with the phase diagrams of 3D
Ising models entering in the Edwards-Anderson type with isotropic quenched bimodal exchange couplings 
either on simple cubic lattice~\cite{hasenbusch_2007,hasenbusch_2008,ceccarelli_2011} or on FCC lattice 
restricted to the FM side (non frustrated)~\cite{thanh-ngo_2014}.
The disorder parameter, the variance $\s$ in our TAD model, is in the above models the probability 
$p$ of anti-ferromagnetic bonds, (equivalently probability of ferromagnetic bonds in the symmetric case of the simple 
cubic lattice). An important difference with the anisotropic bimodal 3D Ising 
models~\cite{papakonstantinou_2013,papakonstantinou_2015} 
is the strong dependence of $T_c$ on the disorder parameter along the PM/SG line in the latter case. 
The phase diagram we get here is also qualitatively comparable to that of the diluted dipolar Ising model
with parallel axes~\cite{alonso_2010,andresen_2014} where the disorder parameter is the site dilution $(1-x)$ or 
equivalently the volume fraction $\phi$ when the latter is drawn in terms of our $T^*\propto\Phi$. Indeed 
doing this, the strong dependence of $T^*_c(\Phi)$ on the PM/FM (or PM/AFM for the cubic lattice) remains and 
instead of the linear dependence of $T_c$ on $\Phi$ found on the PM/SG line, $T^*_c$ is obviously constant.   
In the following we present the details of both the determination of phase separation lines and the characterization
of the nature of the phases. 
 
  %

 \subsection {Ferromagnetic phase}
 \label {res_pm_fm}
\begin {figure}[h!]
  \includegraphics [width = 0.50\textwidth , angle = -90.00]{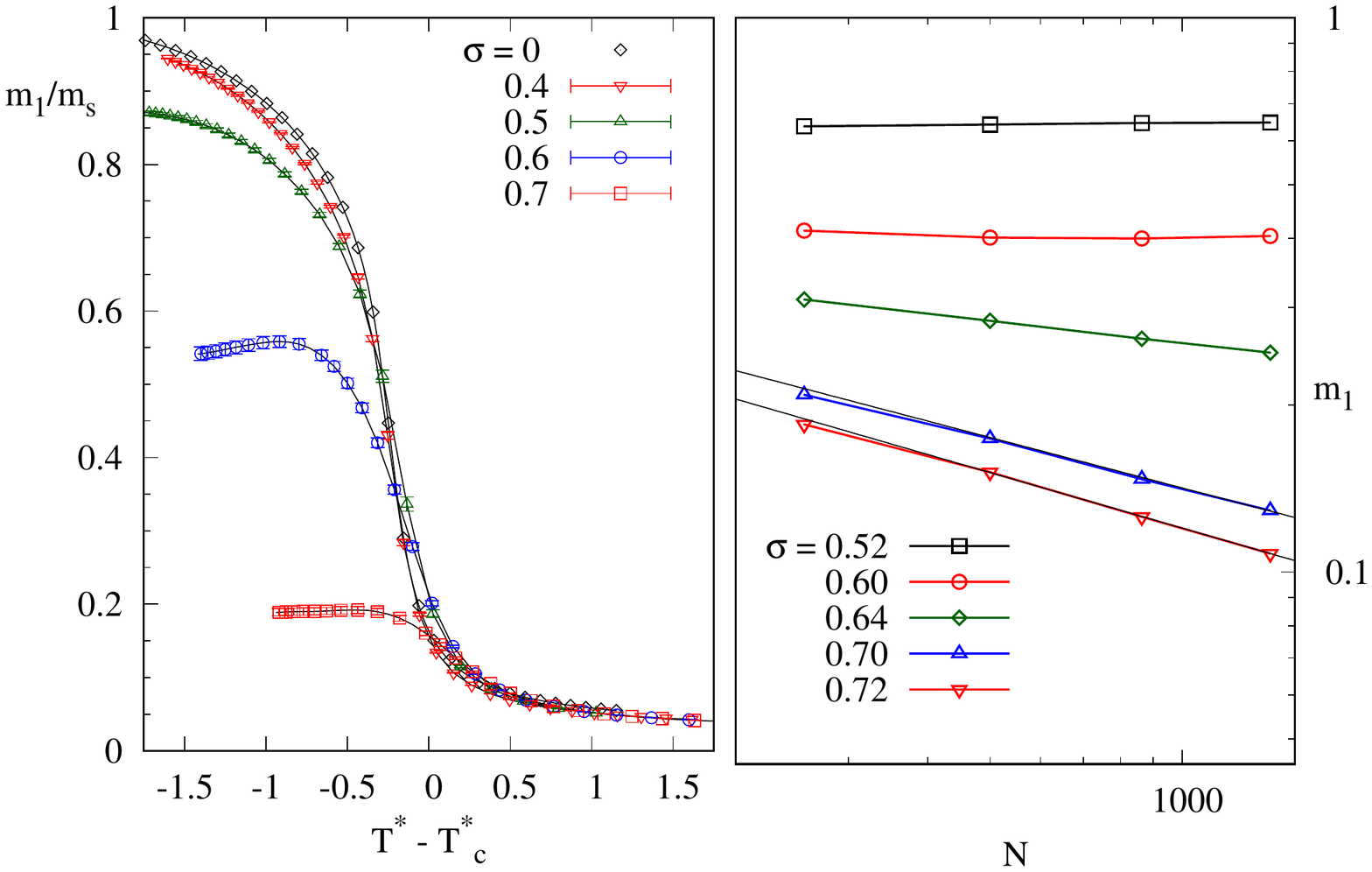}
  \caption {\label {mla_normz_s0-0.7_l7}
   Left: $m_1$ normalized by $m_s=\sum_i|n_{iz}|$ in term of $(T^*-T^*_c)$ for $L=7$ and 
   different values of $\s$ as indicated. The lines result from interpolation by the histogram reweighting 
method~\cite{ferrenberg_1988}.
   Right : log-log plot of $m_1$ in terms on $N$ at $T^*=0.55$ and different values of $\s$. Solid lines are guides to the eye.
    }
  \end {figure}

  We start from a general overview of the evolution of the ferromagnetic (FM) order parameter $m_{\la}(T^*)$ 
with increasing values of the texturation rate $\s$ where the high texturation or ordered state corresponds to $\s=0$. 
To compare the behavior of $m_1$ with respect to $T^*$ with increasing values of $\s$, we take into account that the
maximum value of the spontaneous magnetization $z-$component of a given sample is given by 
$m_s(\s) = \frac{1}{N} \sum_i |n_{iz}|$ which corresponds to a configuration with all the moments up, $s_i{}=+1$,
instead of $m_s=1$ as in the case of the PAD model.
We thus compare, on figure~(\ref{mla_normz_s0-0.7_l7}), $m_1/m_s$ for different values 
of $\s$ and $L=7$ ($N$ = 1372 dipoles). 
As we confirm below, no noticeable change occurs in between $\s=0$ and~0.4. The drastic decrease of $m_1$ and the
related change in the nature of the ordered phase occurs beyond $\s=0.5$. This qualitative picture is made more 
complete by looking at the system size dependence of the polarization at low temperature as can be seen
in figure~(\ref{mla_normz_s0-0.7_l7}) where $m_1$ at $T^*=0.55$ is shown in terms of $N$ for $\s$ ranging from 0.52 
to 0.72 and lead us to conclude that at least for $\s\ge{}0.64$ the polarization vanishes algebraically in the
limit $N\tend\infty$, $m_1\sim{}N^{-p}$ with $p=0.13$ for $\al=0.64$ and 1/3 for $\al=0.72$. 
Accordingly a FM state up to $\s\sim{}0.60$ is expected. 

\begin {figure} [h!]
  \includegraphics [width = 0.50\textwidth , angle = -90.00]{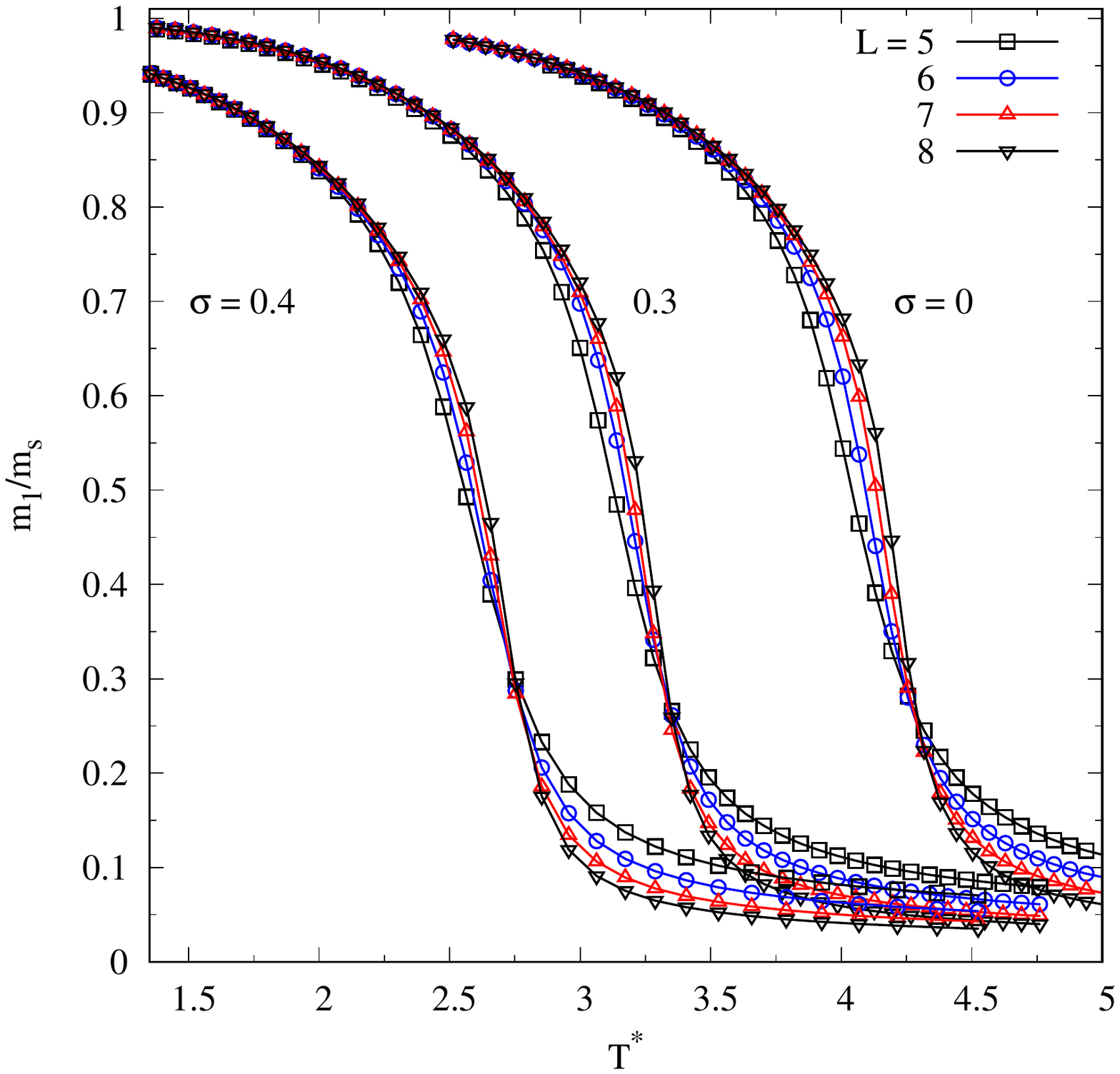}
  \caption {\label {mla_0-0.4}
  $m_1$ normalized by $m_s=\sum_i|n_{iz}|$ in term of $T^*$ for different system sizes ($N=4L^3$)
  and $\s=0$ to 0.4 as indicated. The lines result from from interpolation by the histogram reweighting method.}
  \end {figure}

\begin {figure}[t]
  \includegraphics [width = 0.60\textwidth , angle = -90.00]{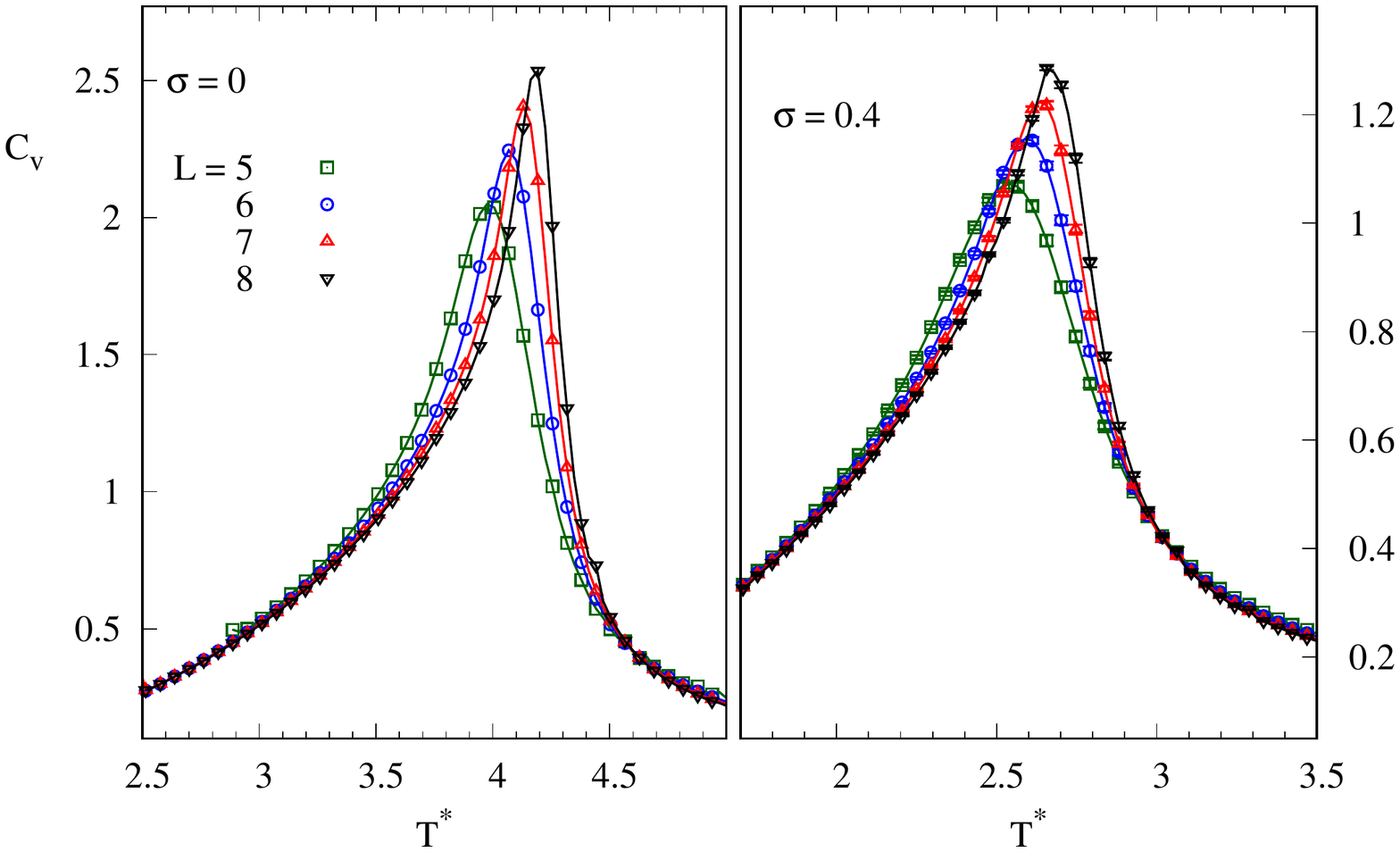}
  \caption {\label {cv_0_0.4}
    Reduced specific heat in unit of $k_b$ in terms of $T^*$ for different lattice sizes.
    Left $\s=0$, right $\s=0.4$.
    }
  \end {figure}

\begin {figure}[b]
  \includegraphics [width = 0.60\textwidth , angle = -90.00]{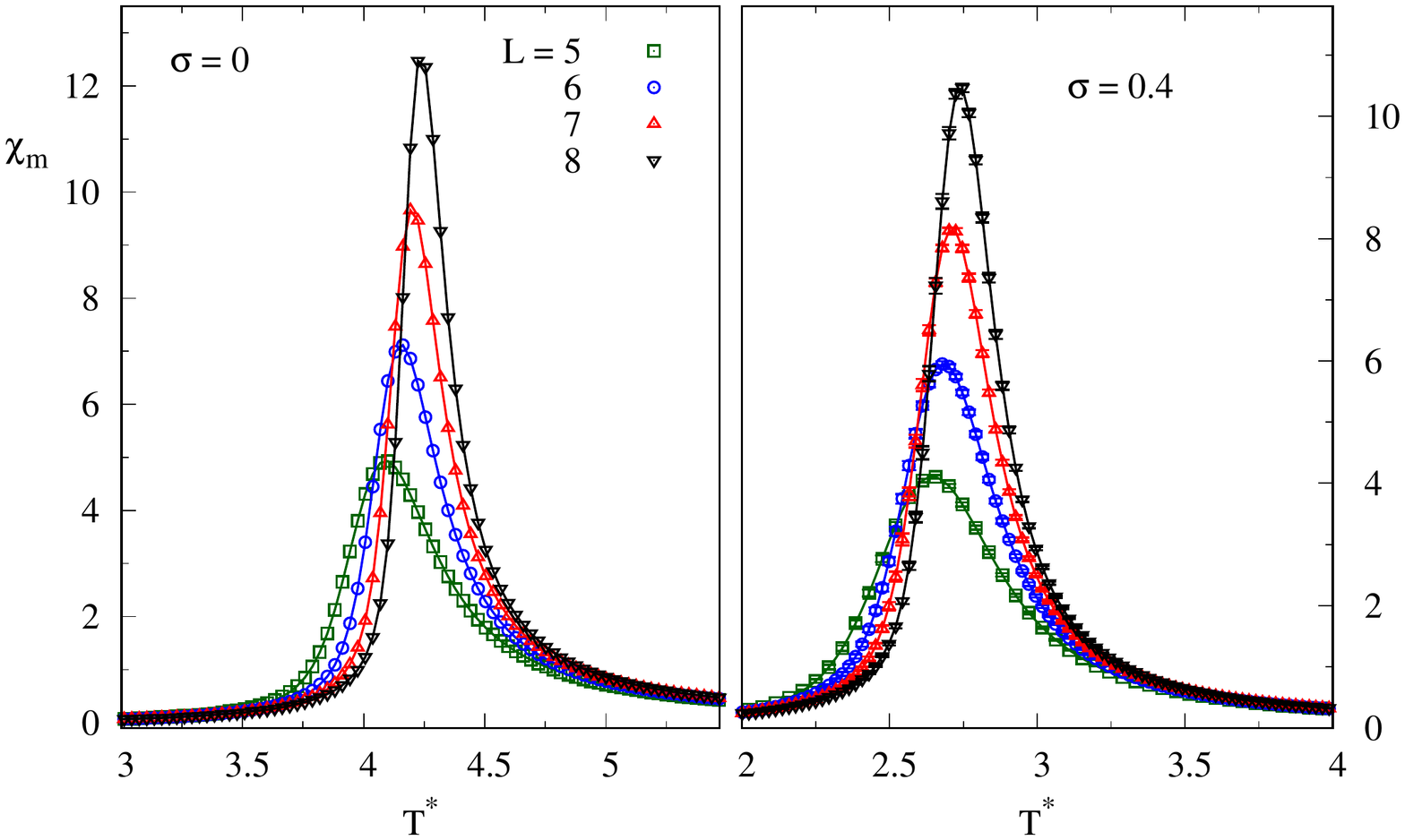}
  \caption {\label {chi_0_0.4}
    Magnetic susceptibility in terms of $T^*$ for different lattice sizes.
    Left $\s=0$, right $\s=0.4$.
   }
  \end {figure}

  Now we focus more precisely on the determination of $T^*_c$ along the PM/FM line.
Let us start by the small values of $\s$, as we know that at $\s=0$, the model orders in a 
well defined FM phase~\cite{fernandez_2000,klopper_2006}. 
As seen above the FM phase can be very well evidenced by the behavior 
of the spontaneous magnetization, or $m_1$ with respect to the temperature, namely a sharp increase of
$m_1$ below the critical temperature $T_c^*$ starting from the nearly vanishing value in the paramagnetic phase 
(see figure(\ref{mla_0-0.4})). 
In figure (\ref{mla_0-0.4}) we compare $m_1/m_s$ for $\s\le{}0.4$ and different system sizes, $L$.
We clearly get a quite similar behavior for all~$\s\le{}0.4$, the main difference being the value of $T_c^*$. 
The main features are first a crossing 
point at a temperature close to $T_c^*$ below (above) which $m_1$ increases (decreases) with $L$, 
the merging at low temperature of the $m_1$ curves for the whole set of $L$-values
meaning that $m_1$ is then system size independent and the saturation to $m_1/m_s(\s)=1$ at $T^*\tend{0}$. 
Only this latter point seems to be not strictly fulfilled at $\s=0.4$. 
%
The specific heat $C_v$ and the susceptibility $\chi_m$ in terms of $T$ are displayed on figures~(\ref{cv_0_0.4},~\ref{chi_0_0.4}).
An important feature is the sharp peak at an effective size dependent $T^*_c(L)$ 
with moreover a clear lambda-shaped curve for $C_v$. These are indicative of a singular behavior of both $C_v$ and
$\chi_m$ with the increase of $L$ characteristic of the second order PM/FM transition. 
The expected scaling behavior of the values taken by $C_v$ and $\chi_m$ at their maxima, say $[C_v]^*$ and $[\chi_m]^*$,
is beyond the scope of the present work. Nevertheless the location of these maxima define $L$--dependent pseudo 
critical temperatures, $T_c^*(L,[C_v]^*)$ and $T_c^*(L,[\chi_m]^*)$.
Besides this qualitative evidence of a PM/FM transition and the $L$-dependent estimation of $T_c$ through 
$T_c^*(L,[C_v]^*)$ and $T_c^*(L,[\chi_m]^*)$, see figure~(\ref{diag_phase_tad}), the precise calculation of $T^*_c$ and 
characterization of the ordered phase is performed through the finite size scaling analysis of the Binder cumulant $B_m$. 

\begin {figure}[h!]
  \includegraphics [width = 0.50\textwidth , angle = -90.00]{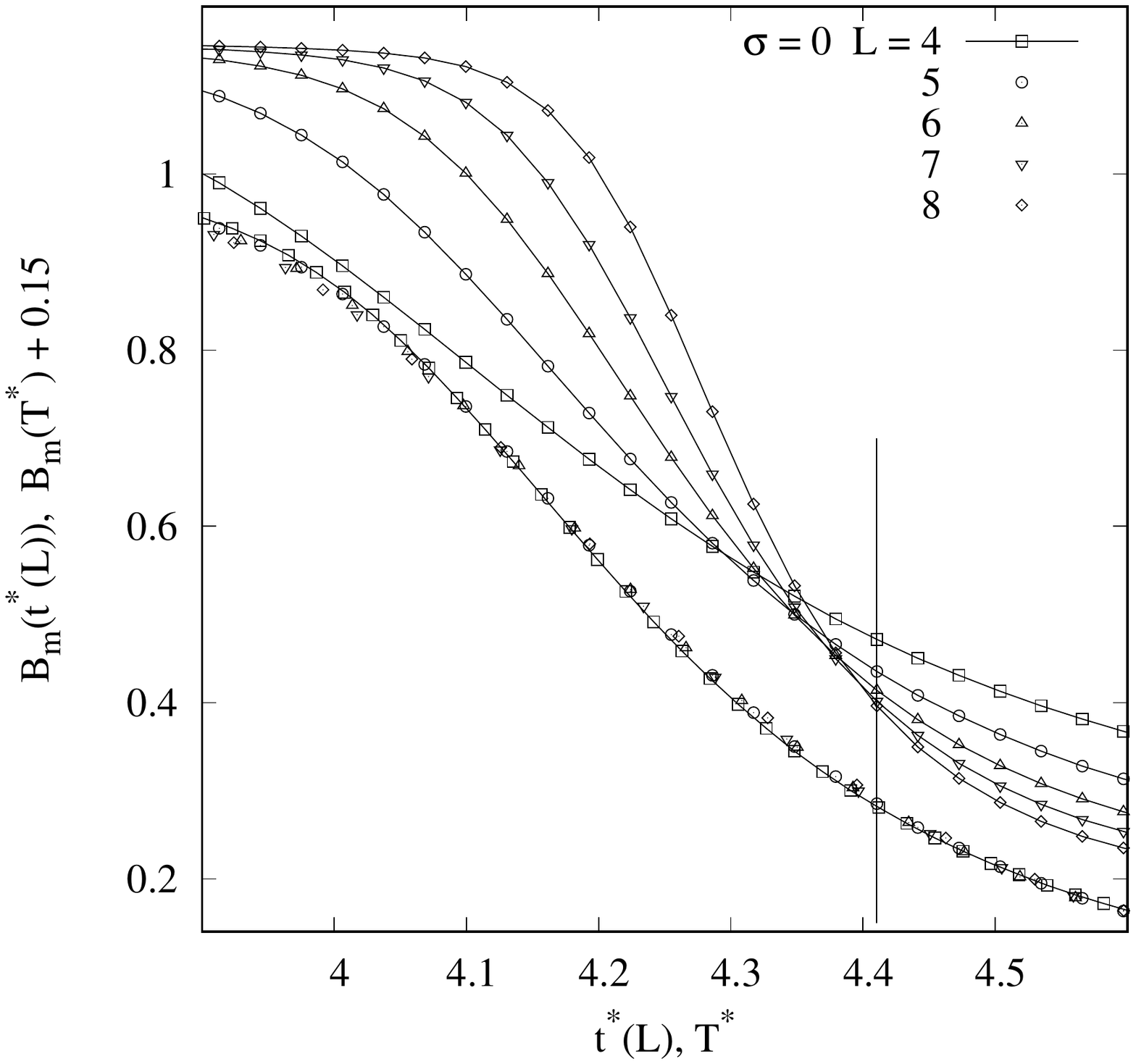}
  \caption {\label {bm_scal-noscal_0}
   Binder cumulant $B_m$ in terms of both $T^*$ and the scaling variable $x_{rg}$ $via$ $t^*(L)$.
   $t^*(L)$ is a $L$-independent linear transformation of $x_{rg}$, $(x_{rg}-a)b+T_c^*$ in order to put $x_{rg}$ and $T^*$
  on the same scale and such that $t^*(L=5)=T^*$ . The error bars are smaller than the symbol scales.}
  \end {figure}

\begin {figure}[h!]
  \includegraphics [width = 0.50\textwidth , angle = -90.00]{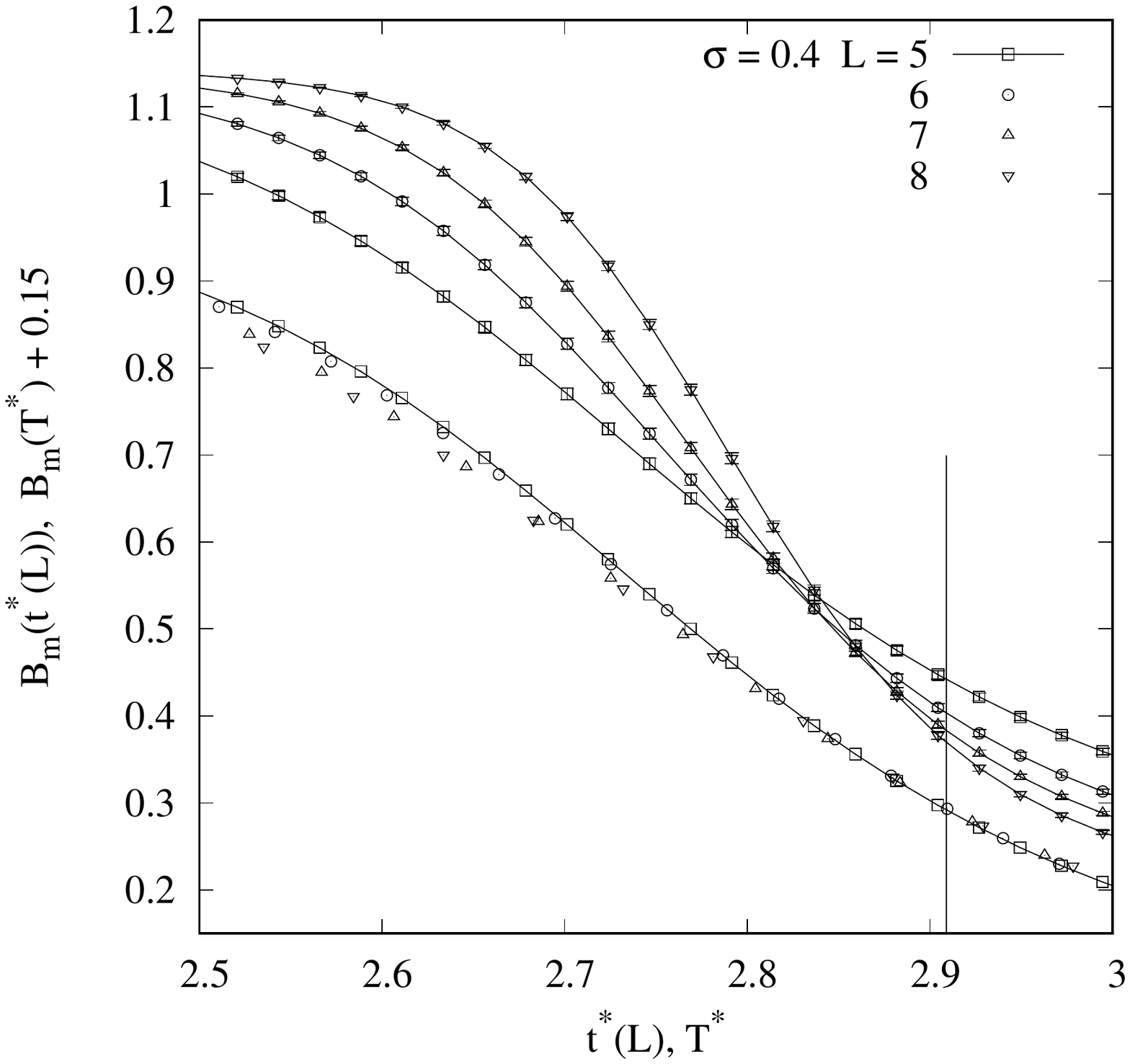}
  \caption {\label {bm_scal-noscal_0.4}
  Same as figure~(\ref{bm_scal-noscal_0}) for $\s=0.4$.
    }
  \end {figure}

\begin {figure}[t]
  \includegraphics [width = 0.50\textwidth , angle =-90.00]{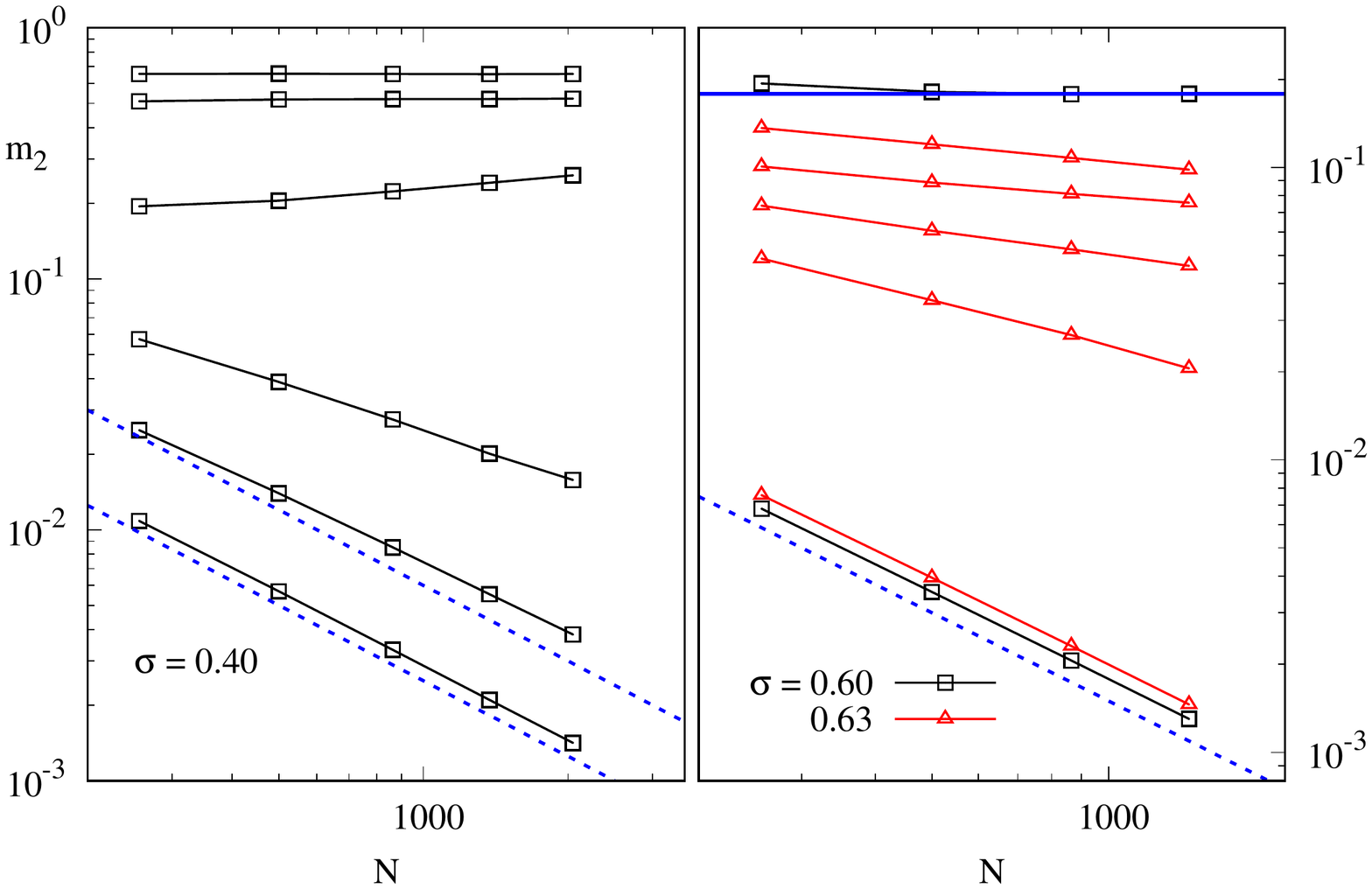}
  \caption {\label {lnm2_0.4-0.62}
  Left : $m_2$ in terms of $N$ for $\s=0.40$ and $T^*=1.31,\;2.00,\;2.56,\;2.96,\;3.41,\;4.52$ from top to bottom.
  $T^*$~=~2.56 and 2.96 bracket the critical temperature  $T^*_c=2.90$. The dashed lines are both proportional 
  to $1/N$, corresponding to the paramagnetic regime, only reached at the largest temperature considered. 
  Right : $m_2$ in terms of $N$ for $\s=0.63$ and $T^*=0.60$, 1.20; 1.40; 1.60 and 3.0 from top to bottom. 
  Also included for comparison the case $\s=0.60$ for the two extreme temperatures, 0.60 and 3.50. The solid 
  horizontal line indicates the values of $m_2$ obtained for $\s=0.60$ at $T^*=0.60$. Solid lines are guides to the eye.
  The dashed lines indicates the $1/N$ decay of the paramagnetic regime. 
    }
  \end {figure}

\begin {figure}[t]
  \includegraphics [width = 0.50\textwidth , angle = -90.00]{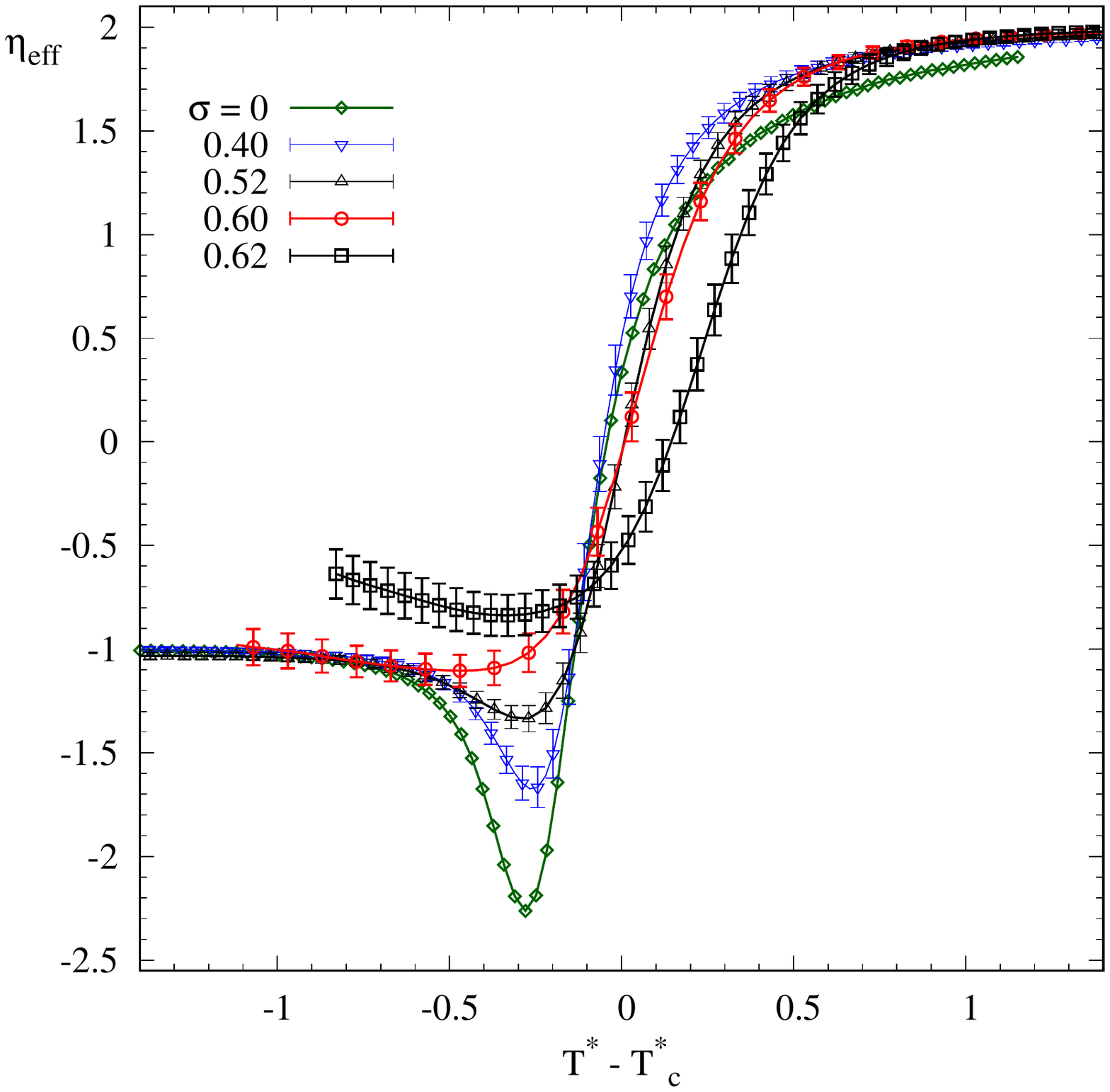}
  \caption {\label {eta_eff_0-0.62}
   Effective correlation function exponent, $\eta_{eff}$ from equation~(\ref{eta}) with 
 $L=(6,7)$
  except for $\s=0.4$ where $(L_1,L_2)=(6,8)$.
    }
  \end {figure}

The system considered in the present work
are too small to provide a determination, or at least an actual check, of the universality class. Instead,
we use the expected scaling behavior as a mean to determine both the nature of the transition and the value
of the corresponding critical temperature, $T_c$.
In the vicinity of the PM/FM transition, it is known that the upper critical dimension of the uniaxial dipolar 
Ising model is $d_u=3$ and as a result the three dimensional dipolar Ising model considered here for $\s=0$ 
and beyond for small values of $\s$ must fall in the mean field regime at marginal dimensionality. 
From the known results of the renormalization group approach~\cite{aharony_1973,gruneberg_2004,klopper_2006},
we have the following relevant scaling relations
  \equab{rg_1}
  B_m &=& f_b(x_{rg}) ~~~ \textrm{with} ~~~ x_{rg} = L^{3/2}\ln^{1/6}(L/L_0)t + v \ln^{-1/2}(L/L_0)
  ~ ; ~~~ t = \frac{T^*}{T^*_c} - 1
  \nonumber \\
  \chi &=& L^{3/2}\ln(L/L_0) f_{\chi} (x_{rg})  
  \auqeb
As in Ref.~\cite{gruneberg_2004}, we first determine $L_0$ in such a way that the maximum value of the scaled 
susceptibility is size independent and then determine both $T_c^*$ and $v$ entering in the definition of 
$x_{rg}$ in such a way that the whole set of $f_b(x_{rg})$ collapse on a single curve. 
We have obtained a quite satisfying collapse of data from (\ref{rg_1}) for $\s\le{}0.45$ showing the FM character 
of the transition. In figure (\ref{bm_scal-noscal_0}) and (\ref{bm_scal-noscal_0.4}) the result of $f_b$ is shown 
in terms of both $T^*$ and the scaling variable $x_{rg}$ for $\s=0$ and 0.4. From these curves, where the 
result of the optimum value of $T_c^*$ is visualized, it is clear that 
the curves $B_m(L,T^*)$ cross around an estimation of $T^*_c$. However all pairs of curves 
do not cross strictly at the same point, since according to equation~(\ref{rg_1}) at $T^*=T^*_c$, 
$B_m=f_b(v\ln^{-1/2}(L/L_0))$ still depends on $L$.
Consequently, the usual way to determine $T_c$ from the crossing point of $B_m$ necessitates an interpolation to 
the $L\tend\infty$ limit. 
The same behavior has been obtained by Klopper  {\it et al.}~\cite{klopper_2006}. 
Moreover we get a good agreement with the latter for the result of $T_c^*$ in the absence 
of disorder, namely $T^*_c(J_{EA=0})=4.459$ in Ref.~\cite{klopper_2006} with our definition of 
$T^*$ , noting that $T^*$ in~\cite{klopper_2006} is related to $\Phi=\pi/6$, compared to our $T^*_c(\s=0)=4.410$. 
Beyond $\s=0.45$, we cannot anymore collapse the set of $B_m$ curves according to equation (\ref{rg_1}), 
but instead from a critical algebraic scaling, $B_m(L,T^*)=f_b(tL^{1/\nu})$. 
The value we get for $\nu$ is to be taken with care. 
Nevertheless it is worth to mention that we get $\nu=0.70$ and 0.693 for 
$\s=0.50$ and 0.52 respectively quite close to that of the randomly diluted Ising model universality class 
($\nu=0.683$)~\cite{ballesteros_1998}. 
From an interpolation of $B_m(L,T^*)$ in terms of $1/L$ for the lowest temperature studied, $T^*=0.55$, 
we still get $B_m(L,T^*)\tend{}1$ in the limit $L\tend\infty$, evidencing a FM long range order (LRO), up to $\s=0.60$.
For $\s\ge{}0.62$ the FM state looses the long range order and transforms in a quasi long range order
(QLRO) FM phase.
This can be deduced from the size dependence of $m_2$, indicative of the integral of the two points
correlation function~\cite{itakura_2003}, displayed on figure~(\ref{lnm2_0.4-0.62}), 
in logarithmic scale for $\s=0.40$ on the one hand and $\s=0.60$ and 0.62 on the other hand. 
Indeed in the former case 
at low temperature $m_2$ is size independent, then for larger values of $T^*$ and $T^*<T^*_c$ $m_2$ increases with $N$,
and finally decreases with $N$ when $T^*>T^*_c$, in agreement with the behavior shown on figure~(\ref{mla_0-0.4}). 
Conversely when
$\s$ increases beyond $\s=0.60$, the behavior in the ordered FM phase is consistent with an algebraic decrease of $m_2$ 
with respect to $N$ whatever $T^*$, as expected in a QLRO FM phase. It is worth mentioning that no qualitative change is 
this decay is observed when $T^*$ gets smaller than $T^*_c$. 
The paramagnetic regime with $m_2\propto{}1/N$ is reached at large $T^*$ whatever the value of $\s$.
%
The rise of the QLRO with the increase of $\s$ can be visualized from the effective exponent $\eta_{eff}$~\cite{itakura_2003},
\equab{eta}
   \eta_{eff} = 2 - D -\frac{ln(m_2(L_2)) - ln(m_2(L_1))}{ln(L_2) - ln(L_1) }
\auqeb
where $D$ is the dimension of space and $L_1,L_2$ two values of  the system size. At $D=3$ the long 
and short range orders correspond to $\eta_{eff}$~=~-1 and 2 respectively.
The result for $\s$ ranging from $\s=0$ to 0.62 is shown on figure~(\ref{eta_eff_0-0.62}) in terms 
of $(T^*-T^*_c)$:  
$\eta_{eff}$ reaches the LRO limit for $\s\le{}0.52$ and deviates from the latter from $\s>0.60$. 
%
  %
Furthermore for $\s\ge{}0.66$, $B_m(T^*,L)$ at different $L$ does not present a crossing point in $T^*$
indicating a lack of FM order suggesting that the system orders in a SG phase (see figure~(\ref{bq_bm_s0.7})).  

 \subsection {Spin-glass phase}
 \label {res_pm_sg}

\begin {figure}[h!]
  \includegraphics [width = 0.50\textwidth , angle = -90.00]{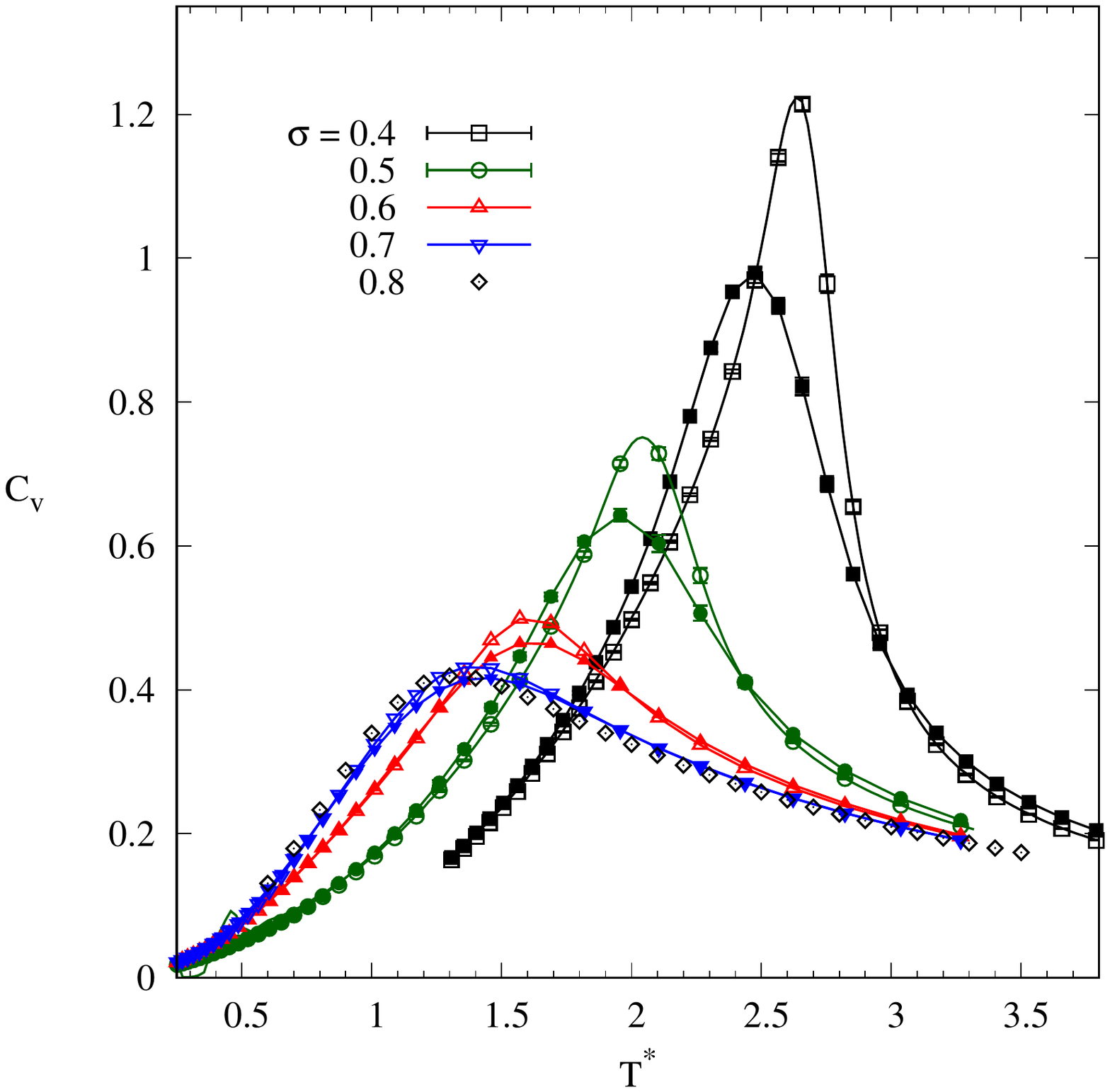}
  \caption {\label {cv_s.4-0.7}
  Heat capacity, $C_v$ in terms of $T^*$ for $\s$ ranging from $\s=0.4$ to 0.7 and in each
  case $L$ = 4 (solid symbols) and 7 (open symbols). For $\s=0.8$, only the $L=7$ case is
  included since it hardly differentiates from the $\s=0.7$ case.
    }
  \end {figure}

  At a qualitative level, the behavior of $C_v$ both in terms of $T^*$ and of the system size $L$ provide 
a simple sketch of the change in the nature of the ordered phase when going from the PM/FM to the PM/SG line. 
This is shown in figure~(\ref{cv_s.4-0.7}), giving a plot of $C_v(T^*)$ for $\s=0.4$ to 0.80 and in each case
for $L=4$ and 7. 
The shape of the $C_v(T^*)$ curve evolves with $\s$ continuously from the lambda-like shape with strong finite 
size effects mentioned in section~\ref{res_pm_fm} for $\s$ ranging between $\s=0$ to 0.4 to a smooth curve with 
no noticeable finite size dependence and thus no singularity expected in the $L\tend\infty$ limit, 
for $\s>{}0.6$. 
This second type of shape for $C_v$
is a well known feature of the PM/SG transition where neither a sharp peak at 
the transition temperature nor size anomaly effect is expected~\cite{ogielski_1985,binder_1986,mydosh_2005} 
with instead a broad peak at a value of $T^*$ slightly larger than $T^*_c$ as is the case for $\s\le{}0.7$ 
in figure~(\ref{cv_s.4-0.7}), a result confirmed from Monte Carlo simulations~\cite{fernandez_2009,thanh-ngo_2014}.
Hence, we deduce that the ordered phase becomes a SG one for $\s$ greater than a threshold value estimated 
in between 0.60 and 0.70 in agreement with the results of section~\ref{res_pm_fm}. 
Moreover, no noticeable change on $C_v$ takes place beyond $\s=0.70$ indicating that the transition temperature 
depends only weakly on $\s$ along the PM/SG line, as confirmed on the phase diagram of 
figure~(\ref{diag_phase_tad}). The precise determination of the PM/SG line is presented below.
  %
  \begin {figure}[h!]
  \includegraphics [width = 0.50\textwidth , angle = -90.00]{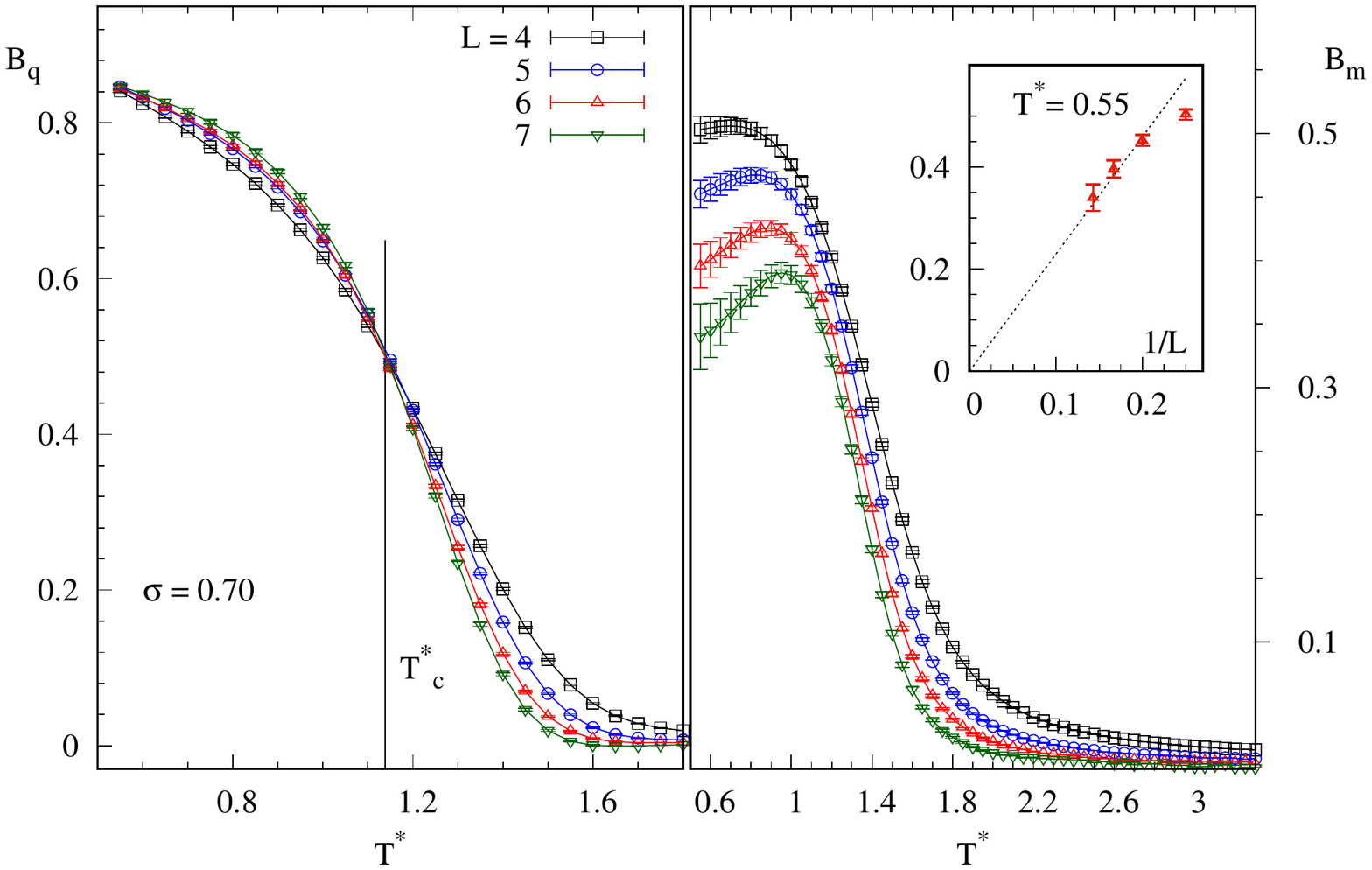}
  \caption {\label {bq_bm_s0.7}
  Binder cumulant relative to the overlap order parameter, $B_q$, left and to the magnetization, $B_m$, right
  for $\s=0.70$. Inset : interpolation in terms of $1/L$ of $B_m(T^*=0.55)$.
     }
  \end {figure}

A reliable localization of the multi-critical point, the common ending point of the PM/FM, PM/SG, and SG/FM lines
is $\s=0.66$.
The criterion to definitely rule out a FM phase with QLRO when $\s>0.66$ is the absence of crossing point 
between the curves $B_m(L,T^*)$ for different $L$ in the whole range of temperature. 
We can then conclude that $B_m$ is a decreasing function of $L$ whatever the temperature.
On the other hand, the Binder cumulant curves related to the overlap order parameter, $B_q$ defined by 
equation~(\ref{bq}) do present a crossing point 
at $T^*_c$, the PM/SG transition temperature. This scenario is displayed in figure~(\ref{bq_bm_s0.7}) for
$\s=0.7$ where we include the dependence of $B_m$ in terms of $1/L$ at the lowest temperature from which
we expect $B_m\tend{}0$ in the limit $L\tend\infty$ at low $T^*$.

We determine the PM/SG transition temperature either from the crossing point of $B_q$ or the collapse of data 
method as used in the PM/FM region  with the algebraic form of scaling for $B_q$ for $\s\ge{}0.7$ in the
vicinity of the scaling region, $t=0$.
The results are summarized in table~(\ref{tab_tc})
and represented on the phase diagram, figure~(\ref{diag_phase_tad}).

\begin {figure}[h!]
  \includegraphics [width = 0.50\textwidth , angle = -90.00]{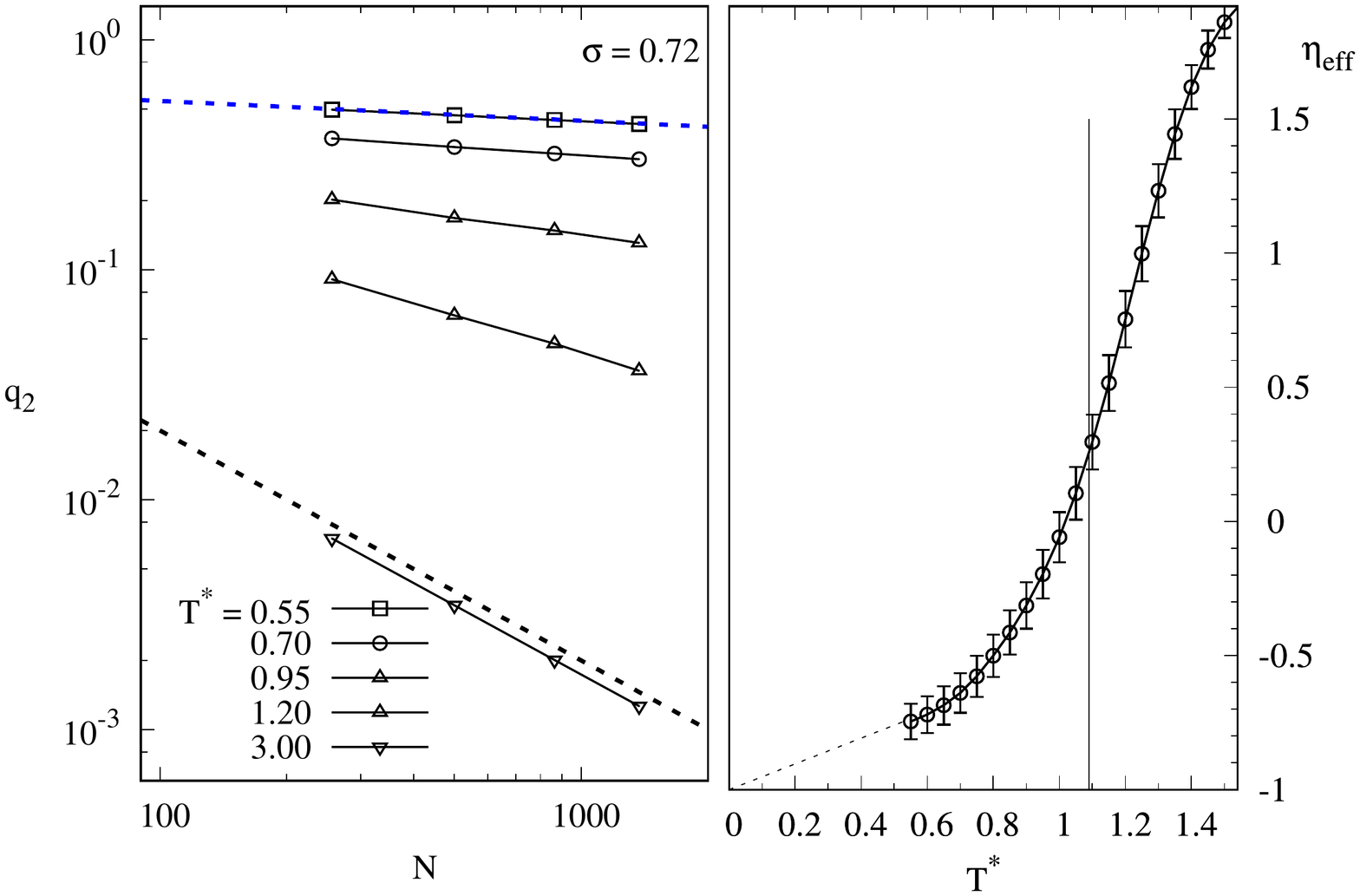}
  \caption {\label {lnq2_s0.72}
  Left :
  $q_2$ in terms on $1/N$ for $\s=0.72$ and $T^*=0.55\;0.70\;0.95\;1.20$ and 3.0.
  Solid lines are guides to the eye.
  The dashed line on top corresponds to the fit of $q_2$ in terms of $1/L$. 
  The dashed line on bottom corresponds to the $1/N$ dependence of the paramagnetic regime.
  Right: 
  $\eta_{eff}$ determined from equation~(\ref{eta}) with $q_2$ in place of $m_2$ and $(L_1,L_2)=(6,7)$. 
  The vertical line indicates $T^*_c$.
     }
  \end {figure}

\begin {figure}[h!]
  \includegraphics [width = 0.50\textwidth , angle = -90.00]{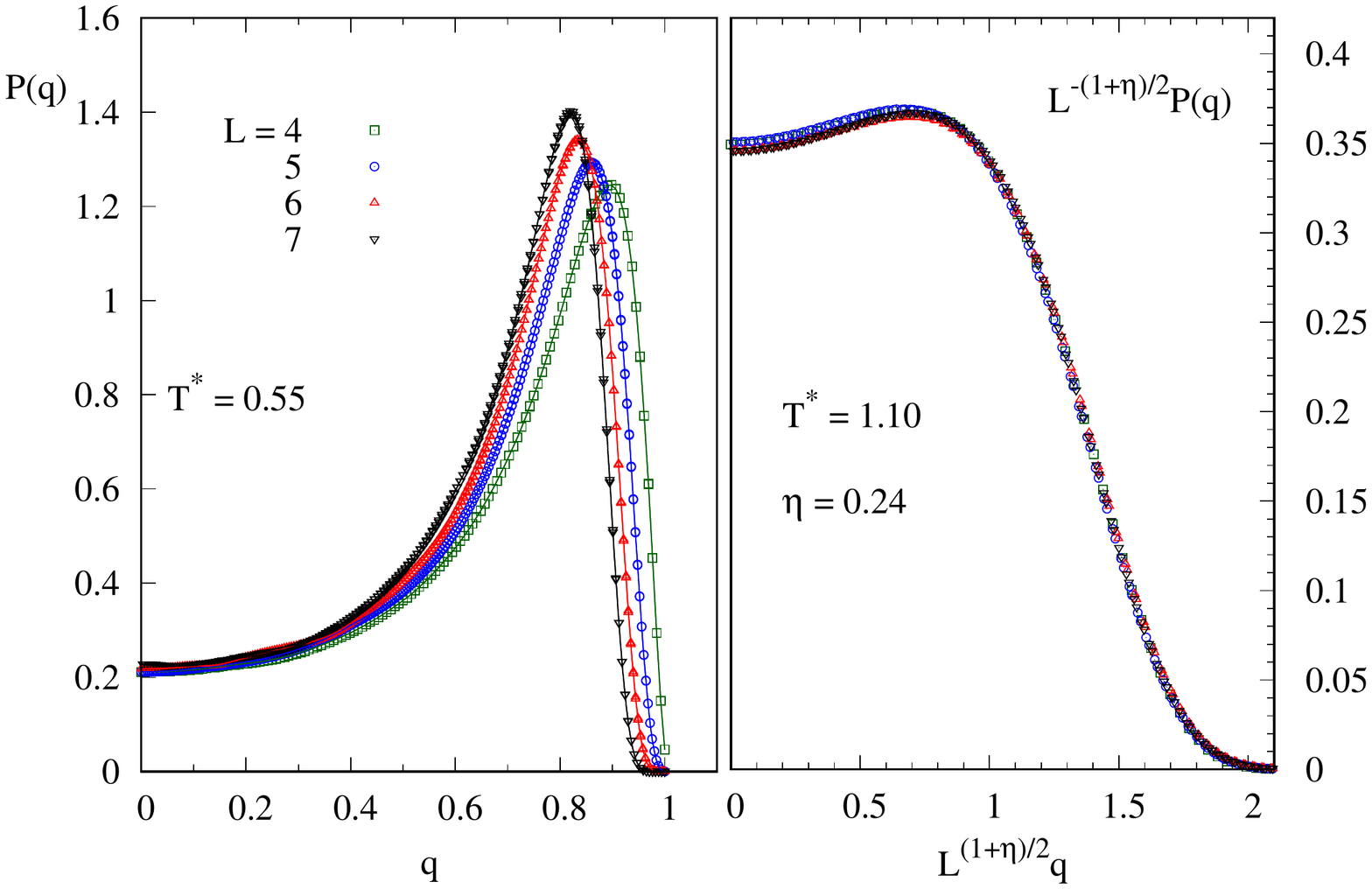}
  \caption {\label {p_q_72}
  Left 
  Probability distribution $P(q)$ of the overlap order parameter for $\s=0.72$ and $T^*=0.55$ for 
  different values of $L$ as indicated.
  Right 
  Scaling plot of $P(q)$ at $T^*=1.10$.
        }
  \end {figure}

  \begin {table}[h!]
  \tab { | c | c | c |}
  \hline
  $\s$ & $T^*_c$ & ordered phase \\
  \hline
  ~ & ~ & \\
  0 	& 4.410(1)   & FM LRO \\
  0.1	& 4.312(1)   & -- \\
  0.2	& 3.976(1)   & -- \\
  0.3	& 3.432(1)   & -- \\
  0.4	& 2.909(1)   & -- \\	
  0.45	& 2.568(1)   & -- \\
  0.52	& 2.11(2)   & -- \\
  0.60	& 1.67(3) & -- \\
  0.62	& 1.57(3) & FM QLRO \\
  ~ & ~ & \\
  \hline
  ~ & ~ & \\
  0.70	& 1.12(4)  & SG \\
  0.72	& 1.09(9)  & -- \\
  0.80  & 1.05(9)  & -- \\
  RAD   & 0.95(9)  & -- \\
  ~ & ~ & \\
  \hline
  \bat
  \caption {\label {tab_tc}
  Values of $T^*_c$ in terms of $\s$. 
  }
  \end {table}
An important point concerns the nature of the SG phase since, in opposite to the result of the FM 
phase where a non ambiguous LRO is obtained for small values of $\s$, the SG phase presents an apparent QLRO, 
that can be deduced from the decay of $q_2$ in terms of $N$ shown in figure~(\ref{lnq2_s0.72}). 
This point has been addressed in the diluted PAD and RAD models~\cite{alonso_2010,alonso_2017} with the conclusion 
of a marginal SG phase characterized by $q_2=0$ in the $N\tend\infty$ limit.
We investigate this problem having in mind as archetypal models the 2D XY model and the 3D bimodal Ising model 
(3D EAI) presenting a KT transition with a line of critical points below $T_c$ and a second order SG transition 
with a SG order below $T_c$ respectively. We characterize the SG phase below $T^*_c$, by using the finite size behavior 
of $B_q(L)$, $q_2$ and the probability distribution $P(q)$ of the overlap order parameter.
$P(q)$ is characterized by two broad peaks located at $\pm{}q_M$ and a plateau with a non vanishing value at $q=0$.
Beside the values of the critical exponents,
the features of the EAI spin-glass phase are~\cite{iniguez_1997,ballesteros_2000} {\it i)} the clear-cut crossing point of 
the $B_q(L)$ curves at $T^*_c$; {\it ii)} the non vanishing value of $q$ in the $L\tend\infty$ limit,
while the marginal or KT phase transition is characterized by~\cite{iniguez_1997,ballesteros_2000} 
{\it i)} the merging of the $B_q(L)$ curves below $T^*_c$; {\it ii)} a vanishing value of $q$ in the $L\tend\infty$ limit.
The point {\it i)} is not strictly discriminating since a crossing behavior has been obtained in the KT 
case~\cite{loison_1999,wysin_2005}, with however a very small splitting of the curves below $T_c$.
A reliable determination of the critical exponents, and in particular $\eta$, is far beyond the scope of this work 
given the computing efforts necessary for this task even on the short range coupling case of the the 3D Ising spin 
glass~\cite{wysin_2005,baity-jesi_2013}.
  The question to answer to discriminate 
  if the SG is a marginal one (of the Kosterlitz-Thouless nature~\cite{kosterlitz_1973}) or a second order one
  like that of the EAI model, is whether $q_2=0$ or $q_2\ne{}0$ in the $N\tend\infty$ limit deep inside the SG phase. 
From our present simulations, we cannot dispel completely a finite limiting value for $q_2$ 
in the $L\tend\infty$ limit for $T^*$ well below $T^*_c$ the lattice sizes considered being too small. 
Nevertheless, by using the whole set of $T^*\le{}T^*_c$, we are lead to conclude that only a fit of $q_2$
according to $q_2\sim{}L^{-(1+\eta)}$, {\it i.e.} with $q_2(\infty)=0$,
 with a $T^*$--dependent value of $\eta$ represent our simulation results.
This result for $\eta$ can be deduced from equation~(\ref{eta}) with $q_2$ in place of $m_2$, see figure~(\ref{lnq2_s0.72}).
It is worth mentioning that the curve $\eta_{eff}$ is close to that expected for the KT 
transition~\cite{itakura_2003,berche_2002}.
Moreover, using the value thus obtained 
for $\eta$ at $T^*_c$ we find that the distribution $P(q)$ satisfies the scaling:
$L^{-(1+\eta)/2}P(q)=f_P(qL^{(1+\eta)/2})$, for instance
at $\s=0.72$ (see figure~(\ref{p_q_72})) a value chosen sufficiently far from the multicritical point. \\
\indent
Then for our model,
the argument in favor of a second order SG phase is the clear-cut crossing point obtained on the set of $B_q(L)$,
from which in principle one is lead to conclude to a well defined transition temperature, $T^*_c$.
However, the SG QLRO phase obtained very likely satisfies $q_2(N\tend\infty)=0$ which corresponds to a marginal SG 
phase as was concluded in~\cite{alonso_2010,alonso_2017}.
%
  %
 
 \subsection {Ferromagnetic Spin-glass line}
 \label {res_fm_sg}

\begin {figure}[h!]
  \includegraphics [width = 0.8\textwidth , angle = -90.00]{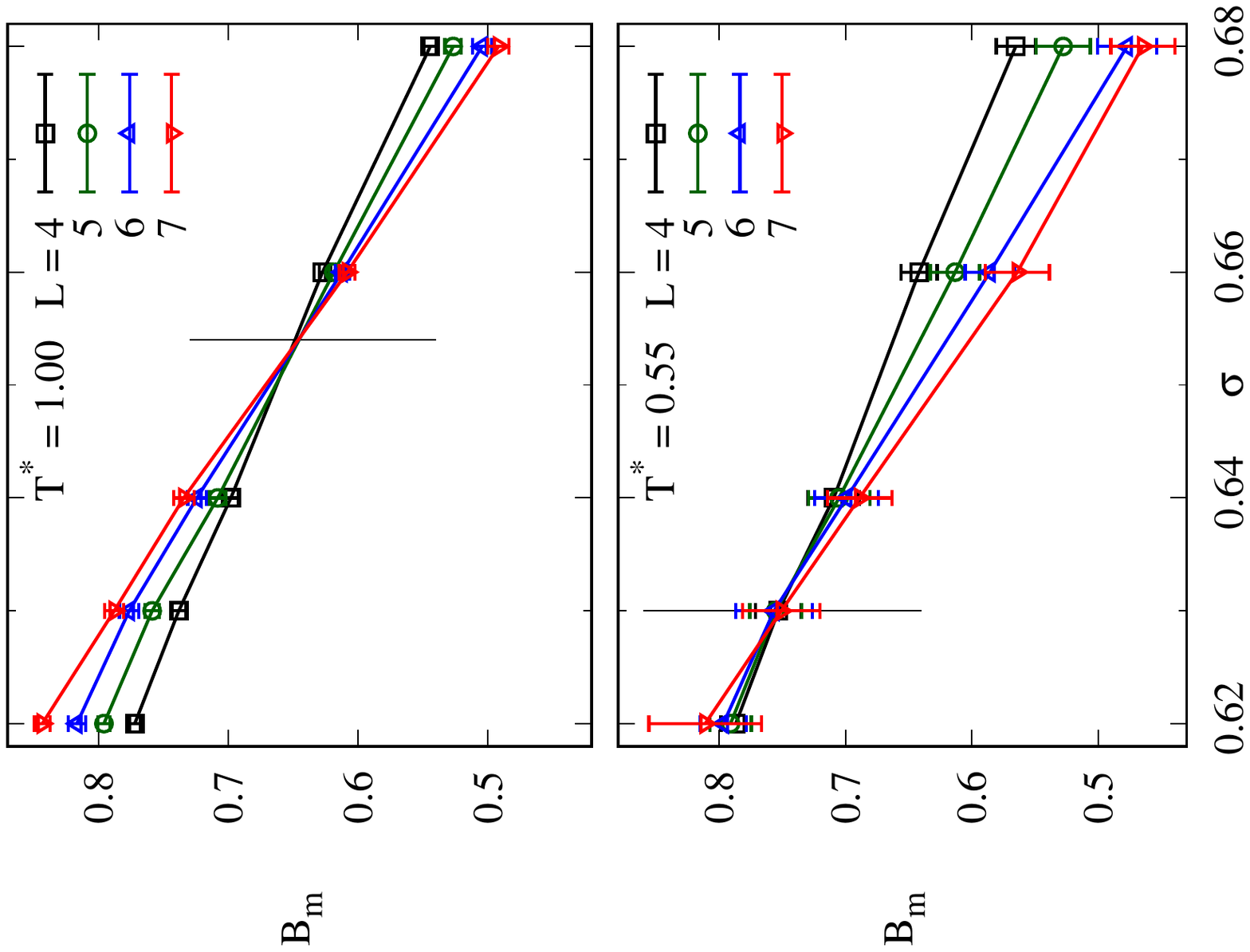}
  \caption {\label {bm_sig_rsg}
  Binder cumulant $B_m$ in terms of $\s$ at fixed temperature as indicated on the figures.
  The vertical lines indicate the corresponding SG/FM transition temperatures.
     }
  \end {figure}

 In this section we focus on the FM/SG line, below the PM ordered transition temperature. We exploit 
the dependence with respect to $L$ of the magnetization Binder parameter, $B_m$ since it is an increasing 
or decreasing function of $L$ in the FM or SG phases respectively. Therefore we consider at fixed values 
of $T^*$ the evolution of $B_m$ in terms of $\s$ as is done in Ref.~\cite{papakonstantinou_2015} in the
case of the anisotropic EA bimodal model.
The first issue to be solved is whether the phase diagram presents reentrant behavior, namely
if the FM/SG line is strictly vertical or not. We have already localized the value of $\s$ at the 
multi-critical point and we are left to determine the slope of the FM/SG line with respect to $T^*$
with sufficient precision to discriminate from the vertical line. 
This is done by using first the two isothermal lines $T^*=0.55$ and $T^*=1.0$. The result for the evolution of
$B_m(L,\s)$ is shown on figure~(\ref{bm_sig_rsg}) together with the crossing points thus determined
on the FM/SG line for these two temperatures. In spite of the numerical uncertainty, we clearly
get the inequality $\s_c(T^*=0.55)<\s_c(T^*=1.0)$ for the critical values of $\s$ on these isotherms.
The other points of the FM/SG line on figure ~(\ref{diag_phase_tad}) are determined in the same way and locate 
to a good approximation on the straight line defined by $\s_c(T^*=0.55)$ and $\s_c(T^*=1.0)$.
\\ \indent
According to the results of section~\ref{res_pm_fm}, the FM/SG line separates the spin-glass and the 
FM one. It seems difficult to precisely locate from our calculations a clear frontier between the true 
LRO and the QLRO FM regions in the phase diagram, but for instance from the behavior of $\eta_{eff}$, 
see figure~(\ref{eta_eff_0-0.62}) or $m_2$ in terms of $N$, figure~(\ref{lnm2_0.4-0.62}), we can roughly 
locate the QLRO FM region between the PM/FM, FM/SG lines and $\s\ge{}0.62$. 
  %
  \section  {Conclusion.}
  \label    {concl}

   In this work we have determined the phase diagram of the textured dipolar Ising
model on a well ordered FCC lattice. The texturation of the Ising axes is
represented by a gaussian like probability of their polar angles characterized
by the variance $\s$ which is then the disorder control parameter in the system.
The determination of the phase diagram is based on a finite size scaling analysis
of the relevant order parameters for either the FM or the SG phase. 
For $\s<0.66$ we get a PM/FM transition, with a LRO in the ferromagnetic phase 
for $\s\le{}0.60$ while in between $\s=0.60$ and the occurrence of the SG phase
a QLRO phase with ferromagnetic character is obtained. 
In the range of system sizes studied here it seems difficult to definitely conclude on the 
nature of the SG phase, 
but nevertheless and despite of the crossing behavior of the curves of the Binder cumulant 
relative to the overlap order parameter $B_q$ we are led to believe that it is a marginal 
spin glass phase. This conclusion results from the behavior of $q_2(N)$ and the resulting 
$T^*$--dependent $\eta_{eff}$ (see figure~(\ref{lnq2_s0.72})). 
Finally a reentrance around the SG/FM line has been obtained.
The phase diagram obtained, figure~(\ref{diag_phase_tad}), looks qualitatively
like those obtained from short range FM/AFM Ising models on simple cubic
lattice or the FM/AFM Ising model on FCC lattice on the FM side. A common
feature is the flat $T^*_c$ PM/SG line in terms of the disorder parameter.
Given the complexity of the computations involved, it is interesting to note that a first
rough estimation of the PM/FM line and the location of the multicritical point can be obtained
from the evolution with $\s$ of the finite size behavior of the specific heat $C_v$ or the 
susceptibility $\chi{}_m$ and the FM order parameter $m_1$. \\ \indent
The comparison with the phase diagram of the DIM with a RCP structure instead of the well ordered FCC
lattice, obtained recently~\cite{alonso_2019}, is important since it present a great similarity when 
displayed in the ($T^*_c,\s$) plane (see figure~(\ref{diag_phase_tad}). Indeed, once the average effect
of the difference in volume fraction is taken into account through our definition of the reduced temperature
the main difference appears to be a shift of the RCP transition lines relative to the FCC ones towards 
smaller values of $\s$ as expected since the RCP structure introduces an additional source of disorder.
Such a similarity is of course not expected for disordered systems at low concentration.
\\ \indent
Finally we note that the range of values of $\s$ where the LRO~FM phase transforms first in the QLRO
and then in the SG phase ($\s\sim{}0.60-0.66$) corresponds to the onset of a non vanishing population
of easy axes with $\Theta\sim\pi/2$ from equation~(\ref{p_theta}).
\FloatBarrier
  \section* {Acknowledgements.}
  \label    {acknow}
  V.~Russier acknowledges fruitful discussion with Drs.~I.~Lisiecki, A.T.~Ngo from the MONARIS, Sorbonne Université and CNRS, 
  J.~Richardi from the LCT, Sorbonne Université and CNRS and S.~Nakamae and C.~Raepsaet,  from CEA Saclay.
  This work was granted an access to the HPC resources of CINES under the allocations 2018-A0040906180 
  and 2019-A0060906180 made by GENCI, CINES, France.  
  We thank the SCBI at University of M\'alaga and IC1 at University of Granada for generous allocations of computer time.
  Work performed under grants
  FIS2017-84256-P (FEDER funds) from the Spanish Ministry and the Agencia Española de Investigación (AEI),
SOMM17/6105/UGR from Consejer\'ia de Conocimiento, Investigación y Universidad, Junta de Andaluc\'ia and 
European Regional Development Fund (ERDF), and ANR-CE08-007 from the ANR French Agency.

  Each author also thanks reciprocal warm welcomes at the University of Màlaga and ICMPE. 
  
%
\FloatBarrier
\bibliography {art_alonso_tad} 

\begin{thebibliography}{60}
\bibitem [{not()}]{note_t*}%
  \BibitemOpen
  \href@noop {} {\ }\bibinfo {note} {Introducing $d_r=d(\Phi_r/\Phi)^{1/3}$
  involving the volume fraction $\Phi$ of the system ($(N/V)\pi{}d^3/6$) and a
  reference value, $\Phi_r$ (for instance the maximum value for hard spheres on
  a FCC lattice), $(1/N)\sum_{i\ne{}j}^{N}(1/(r_{ij}/d_r)^3)=f(N)$ is nearly
  independent of both the underlying structure and $\phi$. As a result
  $(1/N)\sum_{i\ne{}j}^{N}(s_is_j{}D_{ij}/(r_{ij}/d)^3)$ in
  equation~(\ref{edd_1b}) is rewritten as
  ~$(\Phi/\Phi_r)(1/N)\sum_{i\ne{}j}^{N}(s_is_j{}D_{ij}/(r_{ij}/d_r)^3)$.
  Accordingly, $(T_0\ep_d(\Phi/\Phi_r))$ is a convenient measure of the
  temperature for this system from which our definition of $T^*$,
  $T^*=T/(T_0\ep_d(\Phi/\Phi_r))$ follows.}\BibitemShut {Stop}%
\end{thebibliography}
%
\end   {document}